\documentclass[twocolumn,apjfonts]{aastex631}

\usepackage{apjfonts}

\usepackage{graphicx}
\usepackage[suffix=]{epstopdf}
\usepackage{natbib}
\usepackage{amsmath}
\usepackage{url}
\usepackage{xspace}
\usepackage{multirow}
\usepackage{booktabs}

\providecommand{\e}[1]{\ensuremath{\times 10^{#1}}}

\begin{document}

\title{Searching for Stellar Activity Cycles using Flares: The Short and Long Timescale Activity Variations of TIC-272272592}

\correspondingauthor{Tobin M. Wainer}
\email{tobinw@uw.edu}

\author[0000-0001-6320-2230]{Tobin M. Wainer}
\affiliation{Department of Astronomy, University of Washington, Box 351580, Seattle, WA 98195, USA}

\author[0000-0002-0637-835X]{James R. A. Davenport}
\affiliation{Department of Astronomy, University of Washington, Box 351580, Seattle, WA 98195, USA}

\author[0000-0001-5455-6678]{Guadalupe Tovar Mendoza}
\affiliation{Department of Astronomy, University of Washington, Box 351580, Seattle, WA 98195, USA}
\affiliation{Astrobiology Program, University of Washington, Box 351580, Seattle, WA 98195, USA}

\author[0000-0002-9464-8101]{Adina D. Feinstein}
\altaffiliation{NHFP Sagan Fellow}
\affiliation{Laboratory for Atmospheric and Space Physics, University of Colorado Boulder, UCB 600, Boulder, CO 80309}
\affiliation{Department of Physics and Astronomy, Michigan State University, East Lansing, MI 48824, USA}

\author[0000-0001-6147-5761]{Tom Wagg}
\affiliation{Department of Astronomy, University of Washington, Box 351580, Seattle, WA 98195, USA}

\begin{abstract}
We examine 4 years of Kepler 30-min data, and 5 Sectors of TESS 2-min data for the dM3 star KIC-8507979/TIC-272272592. This rapidly rotating (P=1.2 day) star has previously been identified as flare active, with a possible long-term decline in its flare output. Such slow changes in surface magnetic activity are potential indicators of Solar-like activity cycles, which can yield important information about the structure of the stellar dynamo. 
We find that while TIC-272272592 shows evidence for both short and long timescale variations in its flare activity, it is unlikely physically motivated. Only a handful of stars have been subjected to such long baseline point-in-time flare studies, and we urge caution in comparing results between telescopes due to differences in bandpass, signal to noise, and cadence. In this work, we develop an approach to measure variations in the flare frequency distributions over time, which is quantified as a function of the observing baseline. For TIC-272272592, we find a $2.7\sigma$ detection of a Sector which has a flare deficit, therefore indicating the short term variation could be a result of sampling statistics. This quantifiable approach to describing flare rate variation is a powerful new method for measuring the months-to-years changes in surface magnetic activity, and provides important constraints on activity cycles and dynamo models for low mass stars.

\end{abstract}


\section{Introduction}

The 22-year solar magnetic activity cycle is one of the most critical observations to explain with stellar dynamo models \citep[e.g.][]{babcock1961}. Beyond the characteristic timescale, each activity cycle shows dramatic changes in surface properties, including a large scale magnetic polarity reversal, the ``butterfly diagram'' evolution of sunspot emergence from high to low latitudes, and the increase and decrease in occurrence of photospheric and chromospheric activity indicators such as spots, faculae, and flares. While the Sun’s activity cycle has been studied for centuries \citep{eddy1980, usoskin2017}, to understand the formation and evolution of dynamos we must detect activity cycles from other stars. However, activity cycles for stars are notoriously difficult to constrain, as most measurable features of the solar activity cycle are very low amplitude, or require intensive observations to detect from other stars \citep[e.g.,][]{strassmeier_stellar_2005, schrijver_solar_2000}. Activity cycles are therefore only robustly constrained for a few hundred nearby stars, primarily from high resolution spectroscopic monitoring programs over decades \citep[e.g.][]{baliunas1995, egeland2017, baum2022}.

Other observable indicators of stellar activity cycles have been proposed, based on methods used for the Sun. Bolommetric luminosity is known to change over the course of the solar cycle \citep[e.g.,][]{foukal_magnetic_1988}. While individual sunspots can cause large amplitude, short timescale dips in the solar brightness, the Sun is on average brighter during activity maximum due to the presence of faculae \citep[e.g.,][]{wang_modeling_2005}. The signal is very low amplitude, with only a 0.1\% change in ``Total Solar Irradiance'' over the activity cycle \citep{kopp2016}. Missions like Kepler \citep{borucki2010} have been able to detect modulations due to starspots rotating in and out of view, and with careful analysis can detect long-term flux variations that may indicate activity cycles. For example, \citet{montet2017} used Kepler to reveal an evolution from spot to faculae dominated surface activity. However, long-term trends in starspot coverage are difficult to determine for most stars, and spot evolution and differential rotation can confound efforts to estimate changes in starspot behavior \citep{aigrain2015}.

The best characterized stellar activity cycles come from decades long spectroscopic surveys, typically using the Ca II H\&K lines. Surveys like the famous Mount Wilson HK project \citep{duncan_ca_1991} have carried out high resolution spectroscopic monitoring for a few thousand stars over many decades \citep[e.g.,][]{baliunas1995, hall_activity_2007}. While this approach has been successful in detecting chromospheric activity variation (and in some cases bona fide cycles) for nearby stars, it relies on single-object spectroscopy, and is expensive to produce detailed surveys for large samples. 

In this work we advance a relatively new method for characterizing stellar activity cycles, namely stellar flares. As magnetic reconnection events, flares trace the same small scale active regions as e.g. starspots \citep{kowalski2015}. Flares are easily detected in large photometric surveys such as Kepler and the Transiting Exoplanet Survey Satellite \citep[TESS;][]{ricker_transiting_2015}, and their occurrence rates directly correlate with the strength of the stellar magnetic field \citep{davenport2016}. For the Sun, rates of high energy flares directly trace the sunspot activity cycle making them excellent tracers of activity cycles \citep[e.g.][]{veronig_temporal_2002}.

Thanks to the combined long baseline data available from Kepler and now TESS, as well as complimentary ground-based archives \citep[e.g.][]{howard2019}, the long-term evolution of flare rates has started to be explored. \cite{davenport2020} for example showed no significant variation in the flare activity from the benchmark M4 star, GJ 1243. Additionally, \cite{feinstein_evolution_2024} presented eleven young ($<$~300~Myr) stars with annual flare rate variations over five years of TESS observations, but the lack of finer time resolution prevents a robust measurement of activity cycles. 

\begin{figure*}[!ht]
\centering
\includegraphics[height=1.97in]{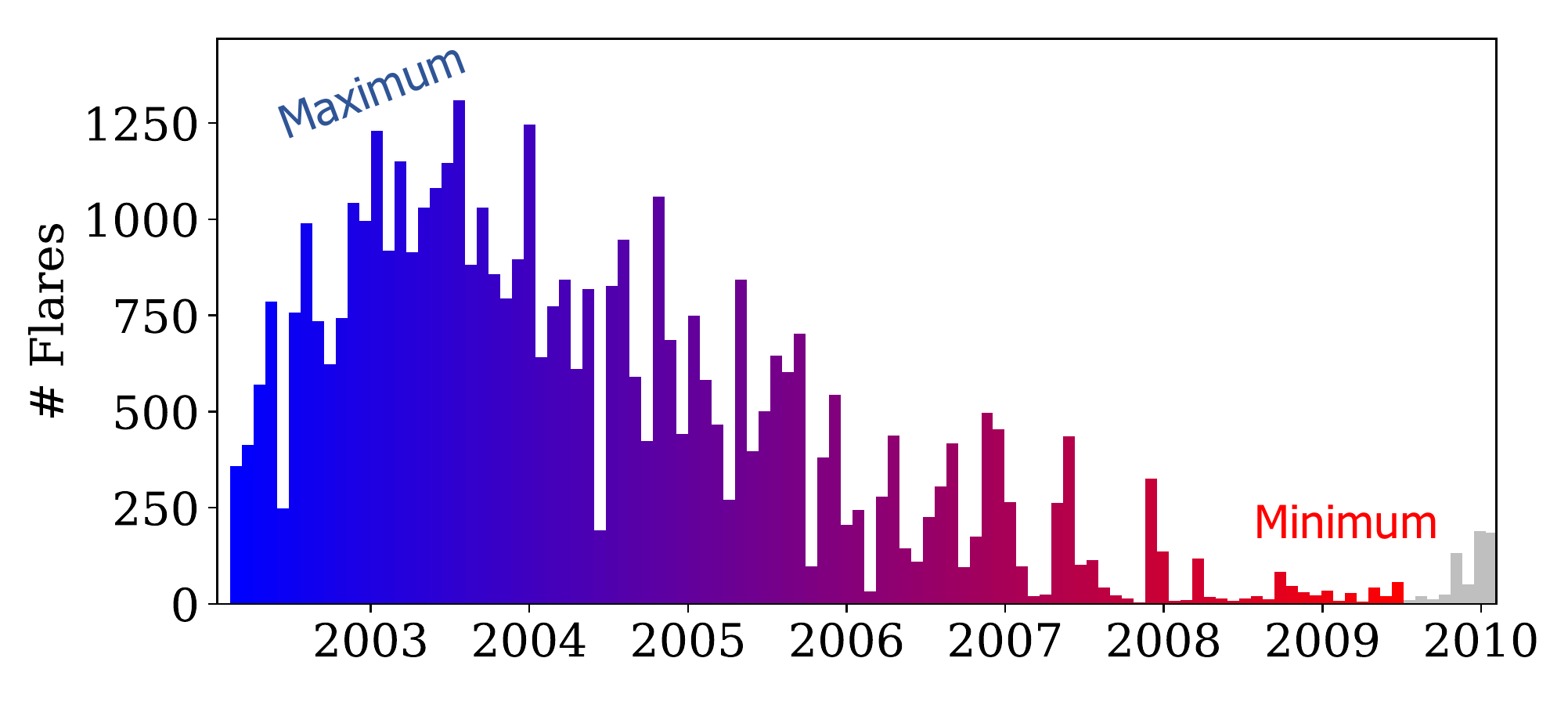}
\includegraphics[height=1.99in]{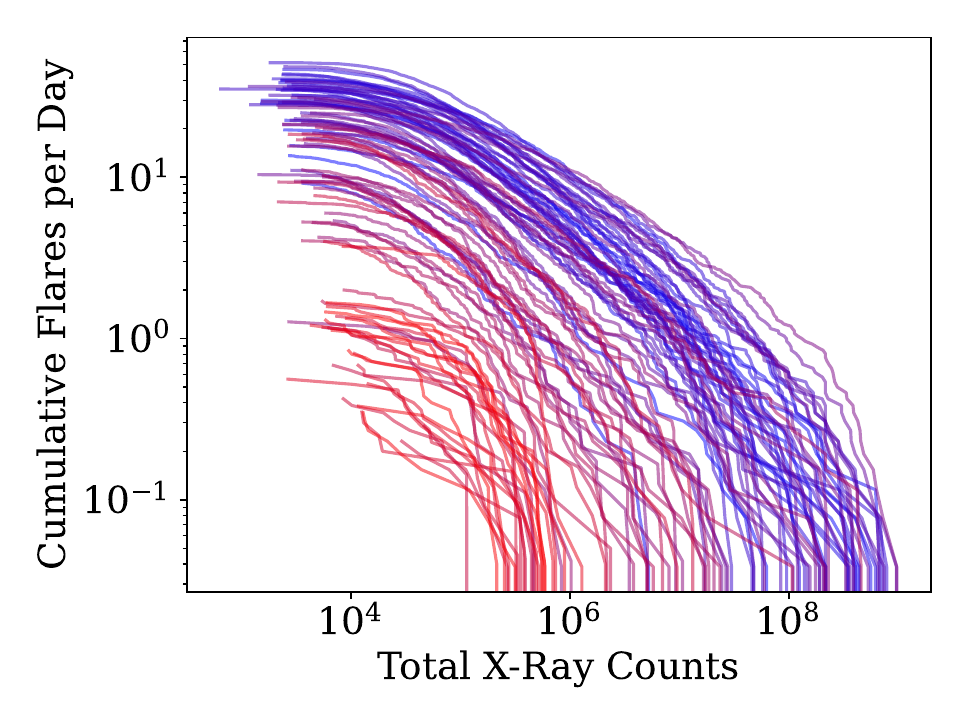}
\caption{Soft X-ray flare data from Solar Cycles 23 from RHESSI. The total number of flares per month is shown (left), which shows a dramatic shift from activity maximum to minimum. The corresponding monthly flare frequency distribution for Cycle 23 (right) is color-coded to match the monthly bins, and demonstrates a nearly two orders of magnitude change in the specific flare rate over the Solar Cycle. Additionally, as highlighted by the cumulative flare frequency distribution, the maximum energy of flares produced by the Sun changes by nearly two orders of magnitude from solar maxima to minima (blue to red).
}
\label{fig:sun}
\end{figure*}

Recently \cite{scoggins2019} identified the M dwarf KIC-8507979/TIC-272272592 (hereafter referred to as TIC-272272592) with known flare activity \citep{davenport2016}, which appeared to have a gradual decline in its flare rate in Kepler data. This star has been re-observed now in 5 Sectors of TESS data over a 3-year baseline, providing an independent examination of both its short (month-to-month) and long-timescale (years) flare behavior. Here we present these new TESS light curves, and a detailed exploration of the flare rate for TIC-272272592.

In \S\ref{sec:sun} we review the motivation for flare rate variation studies, using example data from the Sun. We discuss the Kepler and TESS data in \S\ref{sec:data}, along with how our sample of flares for TIC-272272592 was created. The short- and long-timescale flare rate variations are explored in \S\ref{sec:results}. A thorough examination of the statistical significance of flare rate variations is presented in \S\ref{sec:stats}, and a discussion of upcoming work with TESS is provided in \S\ref{sec:end}.

\section{The Solar Flare Activity Cycle}
\label{sec:sun}

The primary motivation for searching for activity cycles using stellar flares is that we observe similar variations in flare rates and energies on the Sun. During the solar activity maximum, the ``flare index'' (i.e. total daily energy emitted in short timescale flares) increases by a factor of 100-500, rapidly rising over $\sim$2 years, then gradually falling towards activity minimum, as shown in Figure~\ref{fig:sun}. Similarly, the specific flare rate (i.e. number of flare events per day at a given event energy) also varies strongly over the solar cycle. An increase in the specific flare rate of $\sim10\times$ is observed between solar maximum and minimum \citep{veronig_temporal_2002}. This observed flare variation is demonstrated in Figure~\ref{fig:sun}, which shows the number of flares per month over a full 11-year activity Cycle 23 using soft X-ray events from RHESSI \citep{lin2002}. We also show in Figure~\ref{fig:sun} the corresponding per-month flare energy distribution, which shows clear variation of $\sim 2$ orders of magnitude over the activity cycle. The same behavior is seen in H$\alpha$ solar flare data across the activity cycle \citep{yan2012}. 
Importantly, while sunspots show a comparable change in number density over the solar cycle \citep[e.g.][]{morris_solar_2019}, because flares are less degenerate in light curves than starspots, flares are easier to unambiguously count and get accurate energy estimates. Therefore, flares could be more effective tracers of stellar activity cycles than starspot modulations alone. 

The cumulative flare frequency distribution (FFD) shown in Figure \ref{fig:sun} is the standard figure of merit for quantifying flare activity in stars \citep[e.g.][]{lme1976, hilton2011}. The FFD demonstrates the typical power law correlation between flare occurrence rate and event energy, and allows us to quantify the specific flare rate at a given energy \citep{davenport2019}. TESS offers a unique ability to quantify flare rates due to it's continual monitors of stars. While there is no directly analogous data set to TESS available for the sun, the dependence of flare rate on activity cycle seen in the x-ray data suggests it should also be present in the white light of TESS. For solar flares, the specific flare rate varies by more than an order of magnitude across the activity cycle. For stars observed with TESS with a Solar-like activity cycle, this would translate to having 1-2 flares in a given Sector of data in e.g. Cycle 1 if observed at the end of activity minimum, to $\sim$10 flares per day near activity maximum 2 years later (e.g. in Cycle 3). This large amplitude, rapid change in the flare rate should therefore be one of the most unambiguous signals of stellar activity cycles from time series data. 

\section{10 Years of Flare Data}
\label{sec:data}

The M dwarf TIC-272272592 was identified as a rapidly rotating star with a period of P=1.22d \citep{mcquillan2013}, with a $vsin(i) = 14.41$ \citep{jonsson_apogee_2020}, and as a flare-active star using data from the Kepler mission by \citet{davenport2016}. It was further flagged as having potential variation in its flare activity over the 4-year Kepler baseline by \citet{scoggins2019}.
In this Section we introduce new flare data from over 2 years of the TESS mission, as well as a reanalysis of the original Kepler data. In total we have point-in-time activity data spanning nearly 5000 days, allowing us to probe both short term ($\sim$~month), and long term ($\sim$~year) variations in flare rate.

\subsection{TESS Data}
\label{sec:tess_data}
Each Sector of TESS data comprises roughly 27 days of near continuous observations, with 13 Sectors of data obtained across the sky per Cycle (roughly 1-year). The TESS Primary Mission (Cycles 1 and 2) observed nearly 200,000 stars from the TESS Input Catalog \citep[TIC;][]{stassun_tess_2018} with 2-minute cadence using a broad 600-1000 nm passband, centered in the Cousins $I-$band filter.

\begin{figure*}[!t]
\centering
\includegraphics[width=\textwidth]{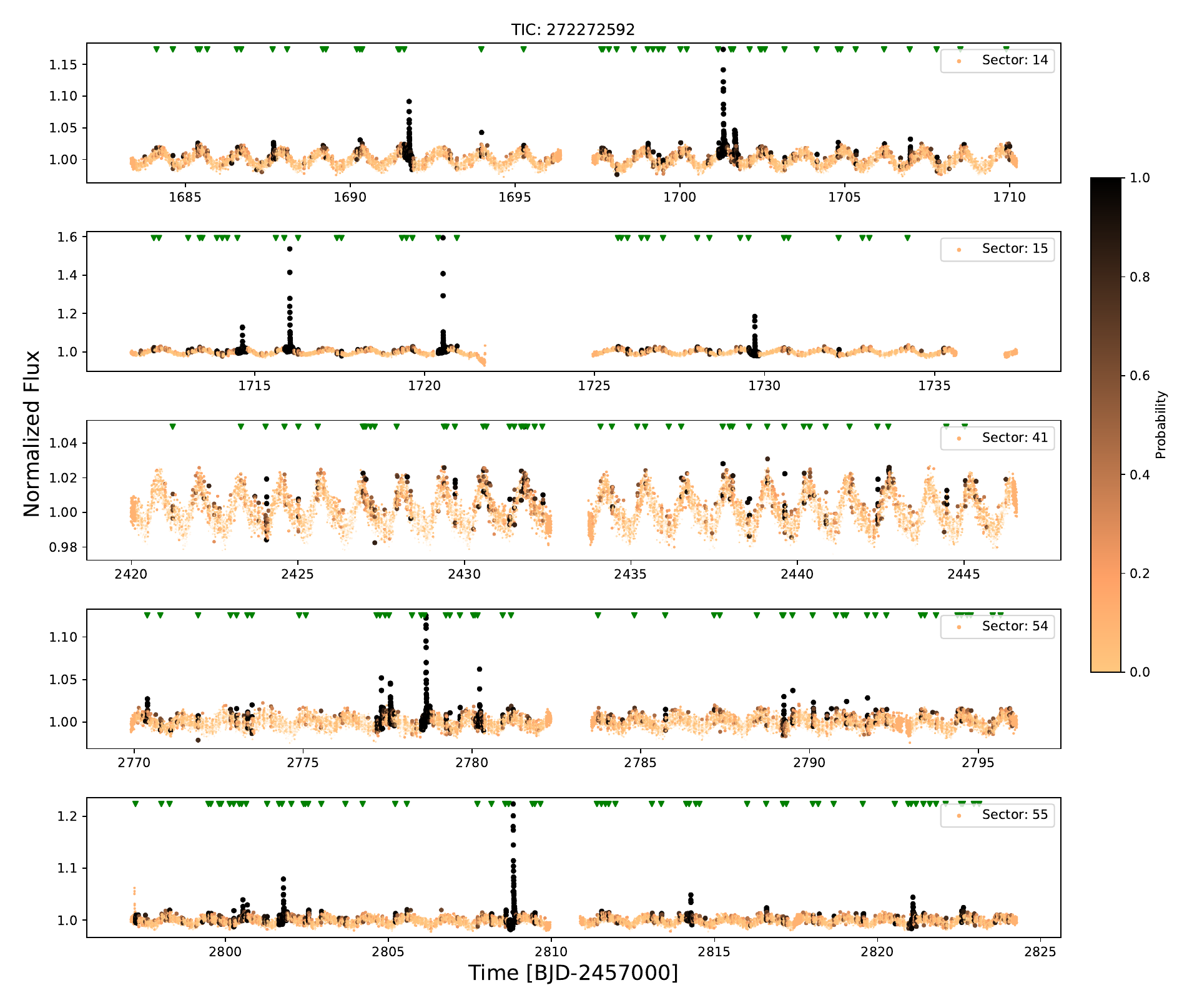}
\caption{Light curves for the five available TESS Sectors of data for the active M dwarf, TIC-272272592. Points are colored and sized by the probability of having a flare, as determined by \texttt{stella}. The location of the start time of each flare identified in our analysis is shown within each panel (green triangles). Noting the y-axis scale for the normalized flux, Sector 41 (center panel) shows a notable lack of large amplitude flares compared with the other TESS Sectors, while all other sectors have flares with amplitudes $A \geq 0.1$.}
\label{fig:tesslc}
\end{figure*}

Staring in 2019 (Sector 14), TESS visited the original Kepler field, re-observing our target star. The Kepler field has now been observed by TESS\footnote{ \dataset[10.17909/fwdt-2x66]{\doi{10.17909/FWDT-2X66}}} five times, with the most recent being in August of 2022 (Sector 55). For this project, we use the light curves from The Science Processing Operations Center \citep[SPOC;][]{jenkins_tess_2016}, utilizing the 2 minute cadence and "PDCSAP" fluxes. We chose these data for the fast cadence, which allows for the detection of shorter duration flares. The five Sectors of TESS data for TIC-272272592 are shown in Figure~\ref{fig:tesslc}. As in the original Kepler data for this star, clear starspot modulations are visible within each TESS Sector, as well as flares. 

 \subsubsection{Flare Identification in TESS Data}
 \label{sec:tess_flare_finding}

To efficiently identify flares in the TESS light curves, we use the convolutional neural network flare finding package \texttt{stella} \citep{feinstein_stella_2020, feinstein_flare_2020}. This code has been trained on a large sample of known flares in the TESS 2-minute cadence data \citep{guenther20}, and generates a probability that each epoch of a given light curve contains a flare based on this training data. Each light curve was ran independently though 10 \texttt{stella} models and averaged as described in \citet{feinstein_evolution_2024}. The light curves in Figure~\ref{fig:tesslc} are colored by this \texttt{stella} probability, and show clear separation between the normal starspot modulations and the impulsive flare events.

While \texttt{stella} will identify and characterize individual flare probabilities from the light curve, we found the identification of flare start and stop times did not reliably work for TIC-272272592. To compensate for this, we used a simple procedure similar to that of \citet{feinstein_evolution_2024} to identify individual flare events from the continuous array of \texttt{stella} probabilities.
We first selected all points having a probability $>30\%$. This first cut was manually selected to eliminate significant contamination from the starspot modulation. Flares were required to contain three consecutive points above this threshold. 
Large amplitude flares often contain complex, multi-peak structure \citep{davenport2014b}. These events can confuse flare identification algorithms, resulting in large flares being broken into smaller events with incorrect energies. To address this, we further grouped together any flare candidates that are within 40 minutes (i.e. nominally 20 epochs) of each other, and consider those points to be a single flare event. This threshold is consistent with complex flare morphology of \citep{zhu_complex_2015}, and was confirmed through a visual inspection of the data, as well as determined to have little impact on the resulting flare sample. The events identified as the start of a flare are marked by the green triangles at the top of each light curve.

While \texttt{stella} is often able to identify low energy flares, automated flare finding tools can still miss small to moderate amplitude flares. We characterized the completeness of \texttt{stella}'s flare identification using an injection and recovery procedure, similar to practices in the literature \citep[e.g.,][]{gao_correcting_2022}. This injection and recovery allows us to directly quantify our flare sample completeness, providing additional accuracy in sample wide measurements in the rest of our analysis.

The injected flare properties in this procedure came from the benchmark, high signal-to-noise sample from \citet{tovar-mendoza2022}, with amplitudes ranging from $0.004\times$ to $0.48\times$ the normalized flux. While these flares are from a different star, they are suitable for exploring the light curve of TIC-272272592 because of the similar stellar properties. Each of the 439 flares from \citet{tovar-mendoza2022} were rendered using the \citet{tovar-mendoza2022} model which can be parameterized by three parameters: flare amplitude, full width at half of the maximum flux, and a center time.
For each TESS Sector, we run the injection of a single simulated flare per trial, rerunning \texttt{stella} in its entirety. Simulated flares were inserted at a random time step in the light curve where existing flares were not located, and injected into all 5 Sectors, a total of 10 times. This process resulted in 21,950 total injection and recovery tests over the 5 Sectors. The flares in the \citet{tovar-mendoza2022} sample were strategically selected to avoid complex flares, and therefore the modeling does not include complex flares. However, complex behavior is mostly present in larger flares \citep{davenport2014b} where \texttt{stella} is most complete, and any affect of complex behavior should not impact the completeness estimate.

\begin{figure}[!t]
\centering
\includegraphics[width=0.47\textwidth]{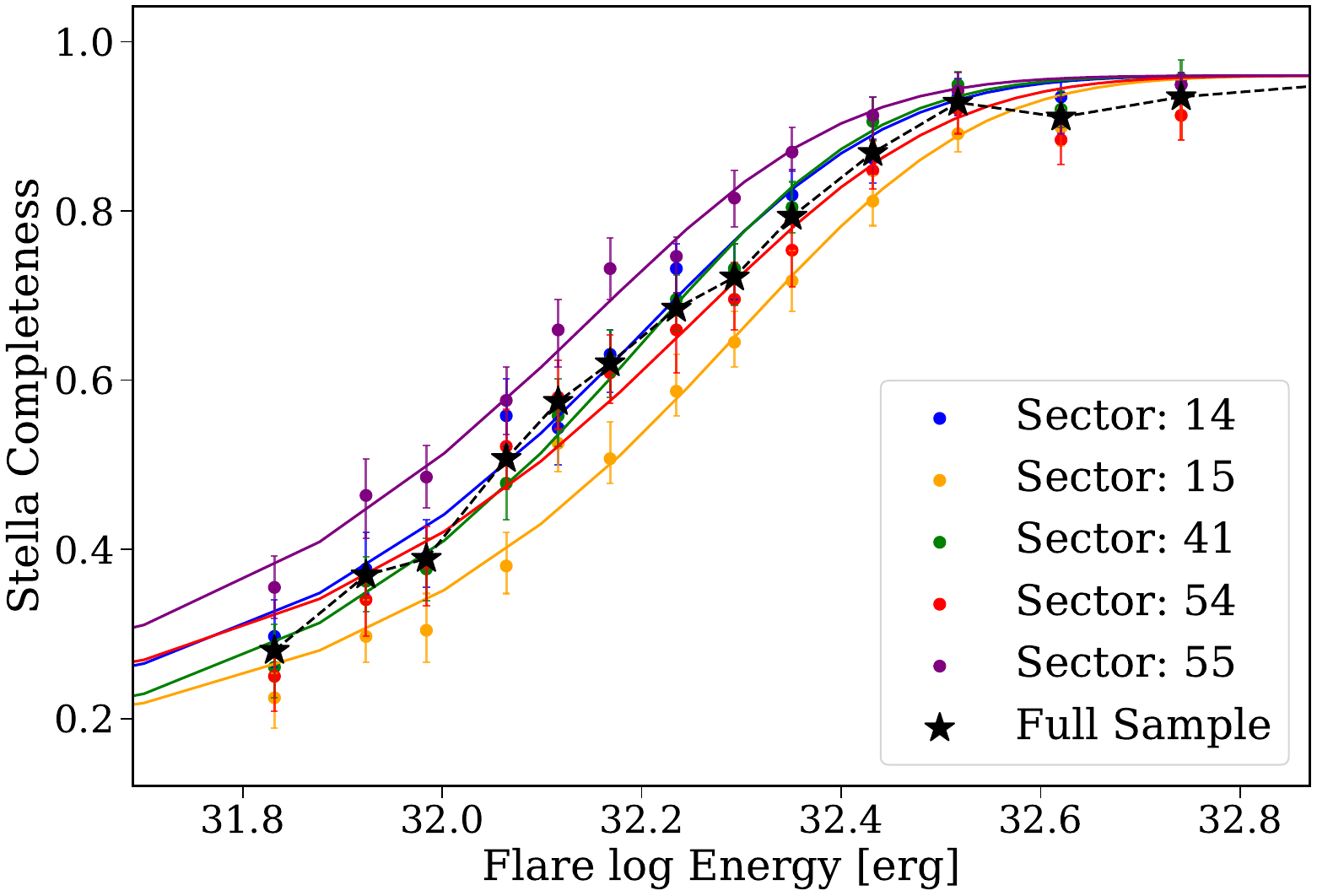}
\caption{\texttt{stella} completeness determined in bins of flare energy from our injection and recovery tests within each TESS Sector, as described in the text (colored points). A logistic function has been fit to each Sector (solid lines).  }
\label{fig:stella_comp}
\end{figure}

From this sample of injected and recovered flares, we measured the completeness as a function of flare event energy.
The flare recovery completeness for this sample is shown
Figure \ref{fig:stella_comp}, both for our entire TESS dataset and also each of the TESS Sectors individually.
We used 15 bins in energy of equal sample size to minimize the completeness uncertainty estimates per Sector, where each bin has 1463 injection and recoveries. We explored increasing the binning and note that the results are not highly dependant on the bin sizes. 
The recovery uncertainty in each bin was determined by performing 1000 bootstrap resamples with replacement of the bin, each resample consisting of 146 injection and recoveries from the energy bin. 
For each completeness curve shown in Figure \ref{fig:stella_comp}, we fit the completeness per bin with a logistic function of the form:
\begin{equation}\label{eq:log}
    \text{completeness} = \frac{0.96}{(1 + e^{-k(x-x_0)})},
\end{equation}
where $x = \log_{10}(E)$ and $E$ is the flare energy in ergs, and $x_0$ and $k$ are the two logistic function parameters we fit over. These curves were fit using a standard nonlinear least squares optimization, and are shown as the solid lines in Figure~\ref{fig:stella_comp}. The fit coefficients, along with the estimated 50\% completeness energies for each Sector are provided in Table \ref{tab:log_pers}.

\begin{deluxetable}{cccc}
\tabletypesize{\small}
\setlength{\tabcolsep}{0.05in}
\tablewidth{0pt}
\tablecaption{Fit Completeness Logistic Function Parameters \label{tab:log_pers}}
\tablehead{\colhead{Sector} & \colhead{k} & \colhead{x0} & \colhead{50\% Completeness}}
\startdata
14 & 0.39 & 4.41 & 32.05 \\
15 & 0.34 & 5.54 & 32.16 \\
41 & 0.43 & 4.62 & 32.08 \\
54 & 0.35 & 4.74 & 32.09 \\
55 & 0.41 & 3.60 & 31.98 \\
\enddata
\tablecomments{Despite being an outlier in the number of energetic flares, Sector 41 is not a outlier in the logistic function parameters, demonstrating that completeness is not the reason why we do not see more flares in Sector 41.}
\end{deluxetable}

We note that while \texttt{stella} does an excellent job overall at recovering moderate to high energy events, it never reaches $100\%$ completeness in this experiment. Consistent with the injection-recovery tests performed in \citet{feinstein_flare_2020}, even the highest energy (and high amplitude) flares have a small chance of being unidentified. This is due to a range of minor issues, including injected flares being too close to other flare events, or within 300 time steps of gaps in the light curve, where \texttt{stella} does not search for flares \citep{feinstein_evolution_2024}. We used a maximum value in our logistic fits of 0.96 as a result, which we consider sufficient for our use in primarily exploring the low energy completeness. Further optimizing the performance of \texttt{stella} and the associated injection and recovery tests is beyond the scope of this paper. 

From these injection and recovery tests, we find that the low energy limit for flare recovery with \texttt{stella} does change between TESS Sectors. We see a shift in the 50\% completeness energy of $\sim$0.25 dex, which is 6 times higher than the typical individual uncertainty in the completeness measurement. These results demonstrate the small variation in data quality between TESS sectors, and illustrate \texttt{stella}'s sensitivity to both instrumental and astrophysical noise.

\subsubsection{Flare Characterization}
\label{sec:flare_characterization}

To quantify the point-in-time flare activity level within a given TESS Sector (or Kepler Quarter) we use the Flare Frequency Distribution (FFD) \citep[e.g.][]{lme1976,hilton2011}, which computes the reverse cumulative distribution of flares as a function of energy. This provides the rate of flares above a given energy.
The FFD is the standard figure of merit for considering flare activity, as it can be compared across observations of differing depths, cadences, or baselines \citep{davenport2019}.

Flare energies are determined based on the integrated relative flux, which is quantified as the equivalent duration \citep{huntwalker2012}. We convert the equivalent duration of each flare to energy by,
\begin{equation}\label{eq: energy from ed}
    \log(\text{Energy}) = \log(ED) + \log(L)  
\end{equation}
where $L$ is the quiescent luminosity of the star given in erg s$^{-1}$, $ED$ is the equivalent duration in units of seconds, and Energy is the flare energy given in erg. For TIC-272272592, using the formula of \citet{davenport2020} matching a Padova \citep{bressan_parsec_2012} isochrone, we derive a quiescent luminosity, $L$, of $10^{31.40}$ erg s$^{-1}$. We note that the intrinsic metric we calculate is equivalent duration, but to compare to literature values, we do this conversion to energy.

Given that stars exhibits quiescent flux variation (i.e. due to star spot modulations) that are not constant throughout a flare event, we determine the baseline integration point from fitting a Gaussian Process (GP) using \texttt{celerite2} \citep{foreman-mackey_scalable_2018}. First, we mask all of the flares in the light curve. Then we fit a two term GP with a short term kernel, and a long term kernel, designed to capture the light curves variability on multiple time scales. Finally, we subtract off the GP fit from the light curve, which allows us to more accurately calculate the flare energies. 

In the TESS data, where we have good constraints on our completeness (Section~\ref{sec:tess_data}), flare rates are corrected for completeness, accounting for flares potentially missed in our process. The correction is based directly on the logistic function fit (Equation~\ref{eq:log}), extending to the $50\%$ completeness limit. The low energy end of the FFD is intrinsically difficult to constrain due to incompleteness in low energy (small amplitude) flares \citep[e.g.,][]{feinstein_evolution_2024}. We explored extending our completeness corrections to lower event energies, but found the resulting FFDs deviated from the expected power law profile, indicating an over-correction for small flares. 

As shown in Figure \ref{fig:stella_comp}, completeness drops off rapidly as a function of flare energy. Thus, pushing to lower completeness limits (e.g. 20\%) only marginally improves the flare sample within each Sector.
Due to the limited statistical benefit, and the increased potential for bias, we adopt the $50\%$ completeness limit for the minimum flare energy in our analysis. For each Sector, the $50\%$ completeness limit is shown in the right panel of Figure~\ref{fig:stella_comp}. Importantly, the $50\%$ completeness limit for Sector 41 is not an outlier, and therefore the lack of large flares observed in this Sector cannot be attributed to \texttt{stella} completeness or unusual noise in the TESS data.

For each TESS sector (and Kepler quarter), we fit the FFD using a power law distribution, which in log-log space takes the form of a line such that,
\begin{equation}\label{eq:beta}
    y= \alpha x +\beta,
\end{equation}
where $y$ is the log of the cumulative flare rate, $x$ is the log of the flare event energy, $\alpha$ is the slope of the power law, and $\beta$ is the corresponding intercept. 

There is evidence in the literature to suggest the FFD of a star is a power law distribution with a fixed slope. While there is not a consensus on the physical driver for the slope of the FFD, with some evidence to suggest the slope of the power law is $\alpha=-1$ regardless of stellar type \citep[e.g.,][]{ilin2020}, there is also evidence that supports the slope of the FFD being dependant on type of star \citep[e.g.,][]{feinstein_testing_2022}. To best determine the power law slope of TIC-272272592, and negate the effects of small number statistics, we combine all TESS sectors into a single FFD and fit the slope $\alpha$, and intercept $\beta$. The fitting was done with \texttt{curve\_fit} function in \texttt{scipy} \citep{scipy}. 

For this combined sample, we find the power law slope $\alpha = -0.85 \pm 0.01$ and $\beta = 27.000 \pm 0.001$. The uncertainties were derived through error weighted least squares fitting, and are underestimated (See Section~\ref{sec:stats}). Throughout the rest of our analysis, we choose to fix the power law slope to this average value of $\alpha = -0.85$ and only fit for $\beta$ in each sector. The slope $\alpha = -0.85$ is in agreement with the population in similar stars in \citet{feinstein_testing_2022}, as well as traditional slope measurements for Sun-like stars \citep{shibayama_superflares_2013}. This decision to fix the slope was also motivated by the noise in individual slope measurements per sector, which impact the resulting $\beta$ measurements, and our ability to compare these metrics effectively. We experimented with fixing the slope to different values (-0.75, -1.0, -1.2, -1.5) and find the resulting analysis of varying $\beta$ values does not appear to be sensitive to the FFD slope.

For each TESS sector, we include in our FFD analysis all flares down to the the corresponding $50\%$ completeness limit  determined through injection and recovery, and assume Poisson uncertainties on the FFD rate. These results will be further discussed in Section~\ref{sec:results}.

\subsection{Kepler Data}

In addition to the five new sectors of TESS data, there are 17 quarters of archival data from  Kepler mission, observed at 30-min cadence.  
As \citet{tovar-mendoza2022} has shown, newer data releases from Kepler can have small changes in the flare sample from earlier surveys for flares in Kepler. Therefore we independently produce a new sample of Kepler flares for TIC-272272592 using the PDCSAP fluxes in the latest available Kepler light curves. 

\begin{figure*}[ht]
    \centering
    \includegraphics[width=0.98\textwidth]{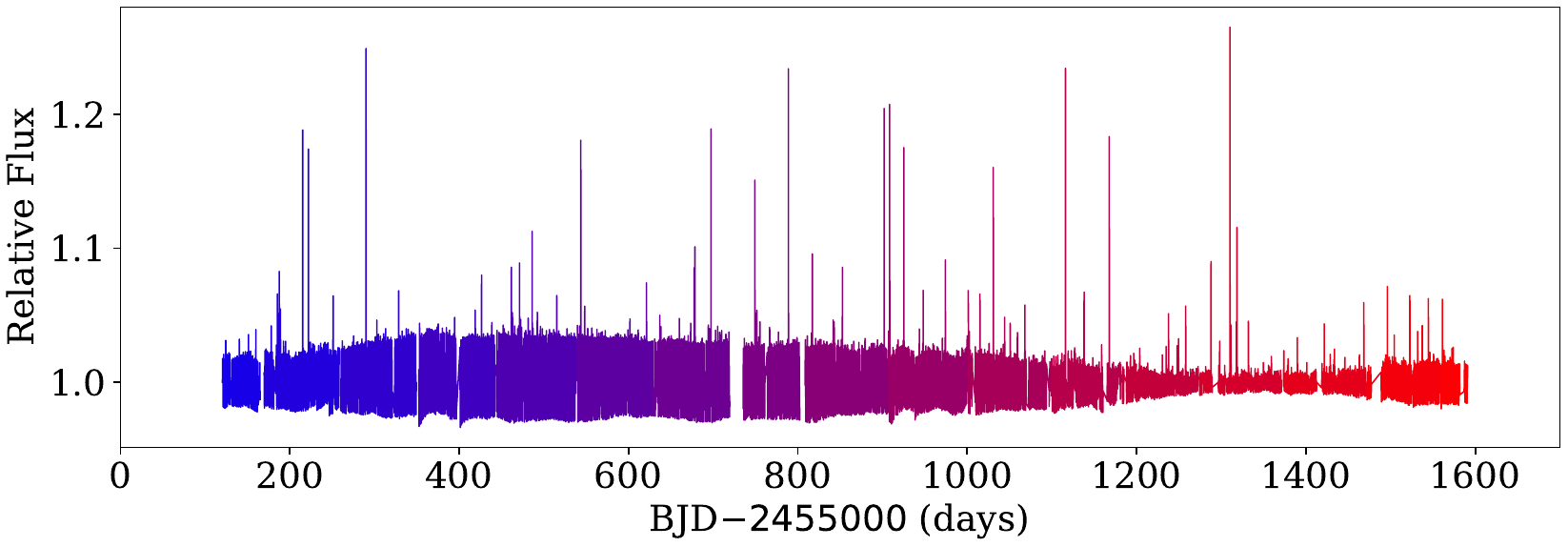}
    \caption{17 quarters of 30-min cadence Kepler data for TIC-272272592, spanning over 1500 days. Each Quarter is colored by time where increasing time corresponds to redder. High amplitude flares occur throughout the light curve. Long timescale changes in the amplitude of rotational variability are also apparent. }
    \label{fig:kepler_lc}
\end{figure*}

As the underlying neural network of \texttt{stella} was trained on 2-minute cadence TESS data, we are unable to use it for the Kepler light curves at present. Instead, for the Kepler analysis we used the \texttt{FBEYE} manual flare toolkit \citep{davenport2014b} to visually inspect every Quarter of Kepler data for this star. While manual flare identification is time consuming, it has been used for many detailed studies in the past, particularly when utilizing low amplitude events \citep{hawley2014}. In total we identified 1251 flares across the 17 Kepler quarters, from November 2009 to November 2013. The previous analysis from \citet{scoggins2019} used a subset of 390 flares identified from a total sample of 1519 flares for TIC-272272592, found by the \texttt{appaloosa} flare finding package \citep{davenport2016}. This subset was chosen to be above the estimated 68\% recovery threshold.  As \citet{davenport2019} note, the \texttt{appaloosa} flare sample particularly struggled with identifying lower energy flares, and in identifying spurious features due to rapid starspot rotations in 30-min cadence data -- i.e. specifically the case for TIC-272272592. 
While our new sample of flares is slightly lower the total yield from \citet{davenport2016}, we were able to remove many spurious events that were identified in the original sample, and improved the flare start and stop times through our manual inspection. 
The combined Kepler light curve is shown in Figure~\ref{fig:kepler_lc}, where each Quarter is normalized by the median flux of the quarter, and a second order polynomial to account for typical Kepler Quarter-to-Quarter systematics.

In the Kepler data, we follow the procedure in Section~\ref{sec:flare_characterization} with the exception of correcting the FFD for completeness since this information is not available in the Kepler sample due to the different mechanism for flare identification. To best accommodate for this difference, and try to remain as consistent as possible across the two samples, we simulate the completeness limit in the TESS data by identifying by eye where the Kepler FFD begins to deviate from a power law, and only fit above this threshold. In the right panel of Figure~\ref{fig:tessffd}, we identified this threshold to be a Energy$=10^{32.8}$ ergs for the Kepler sample, and the best fit lines do not go below this limit. As we did with the TESS data, for the Kepler sample, we keep the slope $\alpha$ fixed to the most likely value of the star, $-0.85$. We also highlight that we calculated the quiescent luminosity in the Kepler band to be $10^{31.06}$ ergs$^{-1}$, roughly 0.4 dex lower than the TESS band, which is incorporated into the flare energy calculation according to Equation~\ref{eq: energy from ed}. 

\section{Results}
\label{sec:results}

In this section, we characterize the short term (month-to-month), and long term (year-to-year) flare rate variability. In Figure~\ref{fig:tesslc}, Sector 41 appears to stand out as having a clear deficit of large amplitude flares, with no flares having amplitudes above 3\% in relative flux.
Long term variation is studied from both the Kepler data alone, and in comparison between Kepler and TESS.

\subsection{Short Term Variation}\label{sec:short_term}

\begin{figure*}[ht]
\centering
\includegraphics[width=0.48\textwidth]{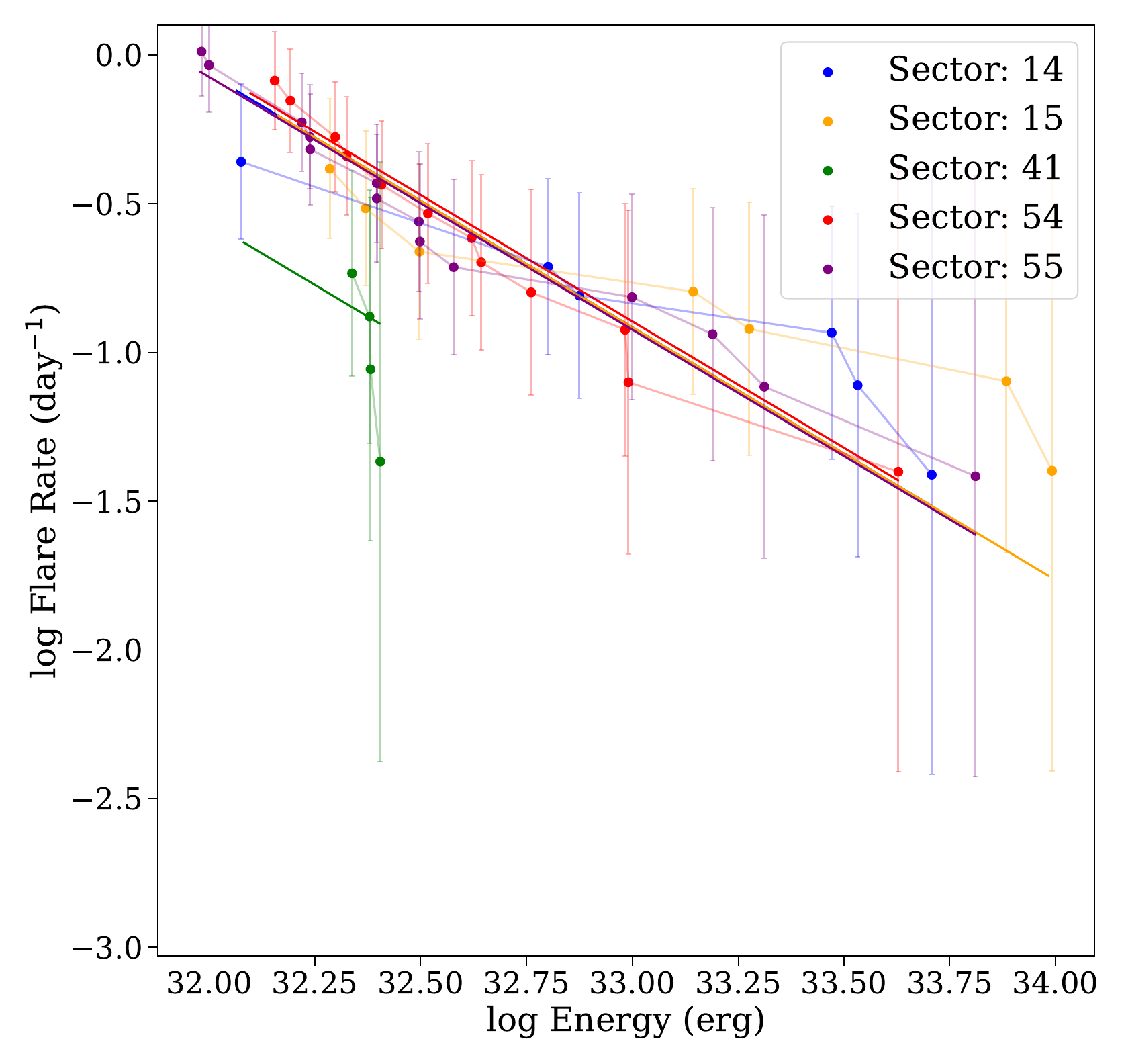}
\includegraphics[width=0.48\textwidth]{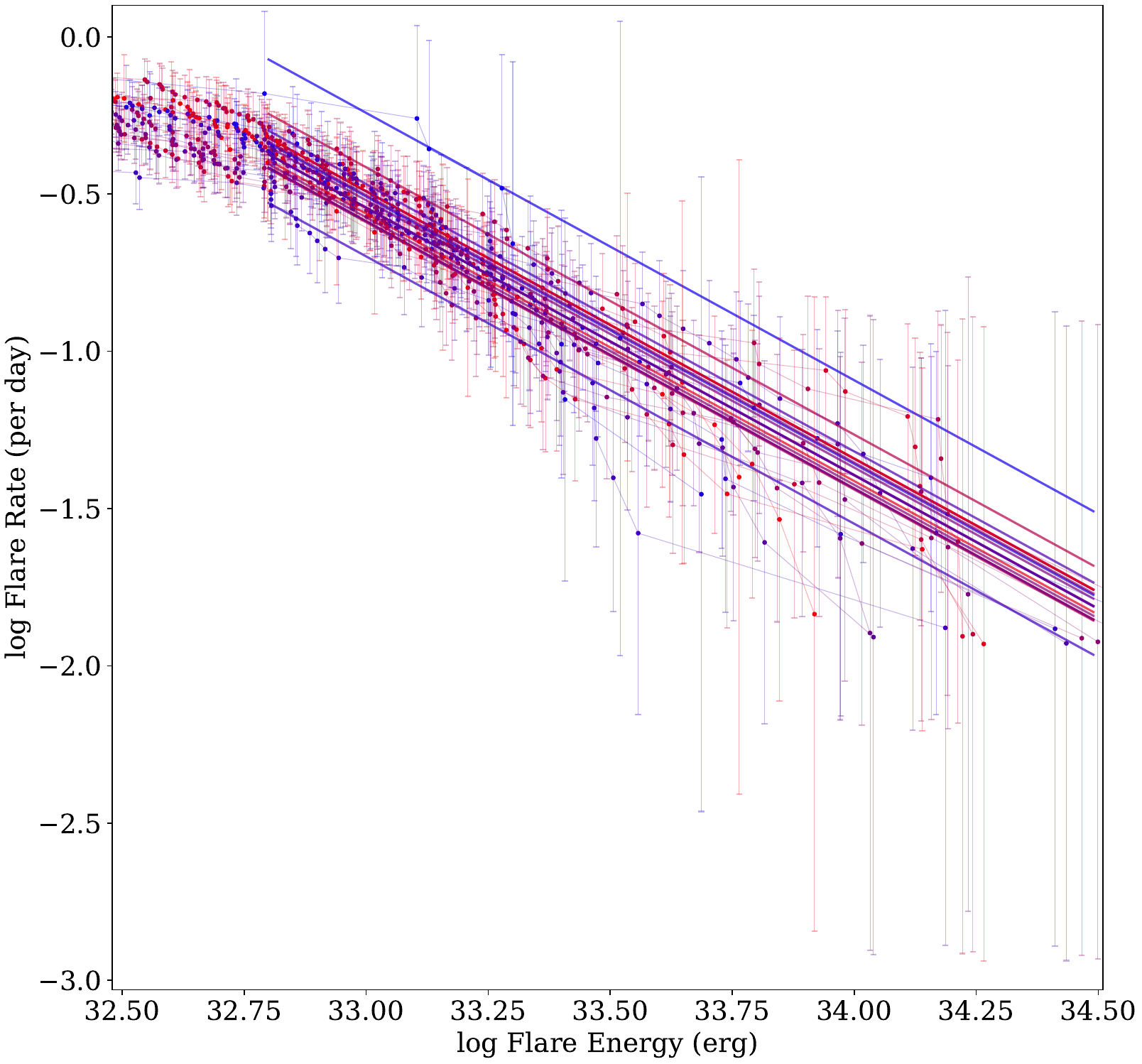}
\caption{Left: Flare frequency distributions (FFD) for each TESS Sector, including completeness corrected flare rates as determined in Section~\ref{sec:tess_data}. Error bars indicate rate uncertainties, where larger energy points have larger errors due to Poisson number statistics. The solid lines represent the best fit lines, with a fixed slope of $\alpha=-0.85$. Sector 41 in green is a clear outlier from the other sectors. 
Right: quarter by-quarter FFD for flares in the Kepler sample. The solid lines represent the best fit lines, with a fixed slope of $\alpha=-0.85$. Colors map to the time axis in Figure~\ref{fig:kepler_lc}, where red is later quarters. It is of interest to note that while the energy ranges are similar between the two telescopes, the energies themselves are not directly analogous across telescopes due to band pass differences.} 
\label{fig:tessffd}
\end{figure*}

The primary results of this work are shown in Figure~\ref{fig:both_t_and_k}, where we present the best fit intercept values $\beta$, for each TESS Sector's and Kepler quarter's FFD. As seen in this Figure, flare activity in TESS Sectors 14, 15, 54, and 55 are all extremely similar, with fit $\beta$ values of 27.13, 27.13, 27.15 and 27.13 respectively (as also visualized in the FFDs of Figure~\ref{fig:tessffd}). However, Sector 41 again appears to be an outlier, with a maximum $\log_{10}(E)$ of only 32.40, and a fit intercept of 26.64, over a half a dex below the mean flare activity levels in the other TESS Sectors. While we are unable to confirm that the lack of large flares is due to an unlikely telescope or pipeline systematics (i.e., anomalous cosmic ray rejection), the star spot rotation in this quarter remains preserved, and suggests that the processing is not responsible. Another explanation is the measurement is biased by the observing window, and we will further discuss the statistical significance of this deviation in Section~\ref{sec:stats}.

In the Kepler data we choose to remove Quarter 0 from our analysis, as it only contained 17 days of data, and therefore has significantly higher uncertainties than the rest of the Kepler data, as we will discuss in Section~\ref{sec:stats}. Each Quarter was fit with the same power law FFD model, assuming a fixed $\alpha$, as shown in Figure \ref{fig:tessffd}. Our derived $\beta$ values for the remaining 16 quarters are also visualized alongside the TESS values in Figure~\ref{fig:both_t_and_k}, with point colors corresponding to those shown in Figure \ref{fig:kepler_lc}. The Kepler flare activity level is remarkably flat over the 4 years of data analyzed, and we do not find the same decline in flare activity over time suggested by \citet{scoggins2019}. 
This discrepancy with earlier analysis for TIC-272272592 is likely due to incorrect flare energy estimates, especially for long duration and high amplitude events 
from the original \citet{davenport2016} algorithm, especially from long cadence (30-min) data, and in the presence of changing starspot amplitudes (e.g., see Figure \ref{fig:kepler_lc}).
We note that the FFDs analyzed by \citet{davenport2016} and by extension \citet{scoggins2019} showed a break in the FFD for the highest energy flares, which we find is due to incorrectly estimating the starspot behavior ``under''  the flares, resulting in unreliable flare activity estimates over time. When we correct the flare start and stop times, and the starspot model from the original \citet{davenport2016} sample, we find no long duration decline is seen, consistent with our manually identified sample above.

\subsection{Long Term Variation} \label{sec:long_term_var}

\begin{figure*}[ht!]
\centering
\includegraphics[width=0.48\textwidth]{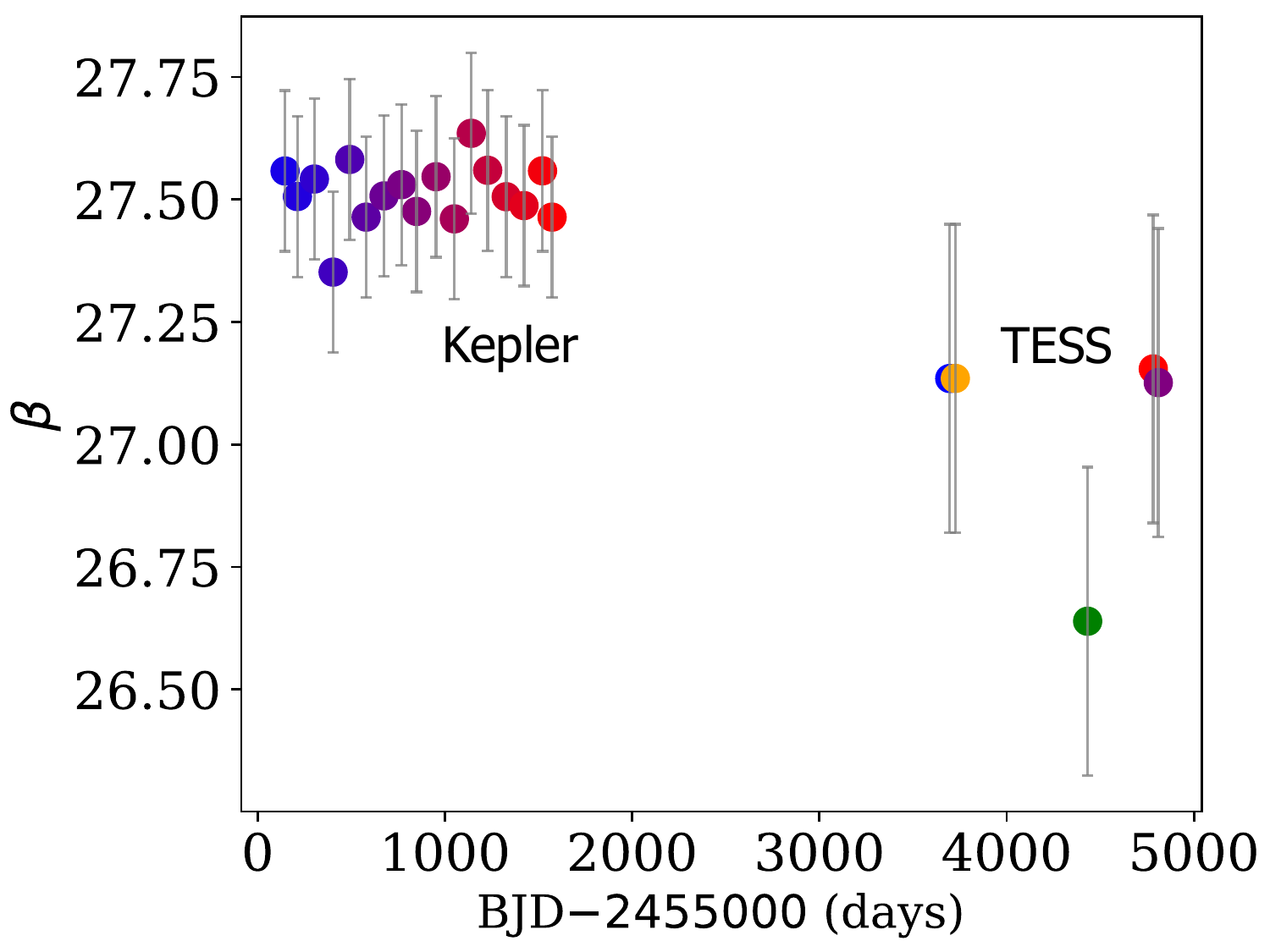}
\includegraphics[width=0.49\textwidth]{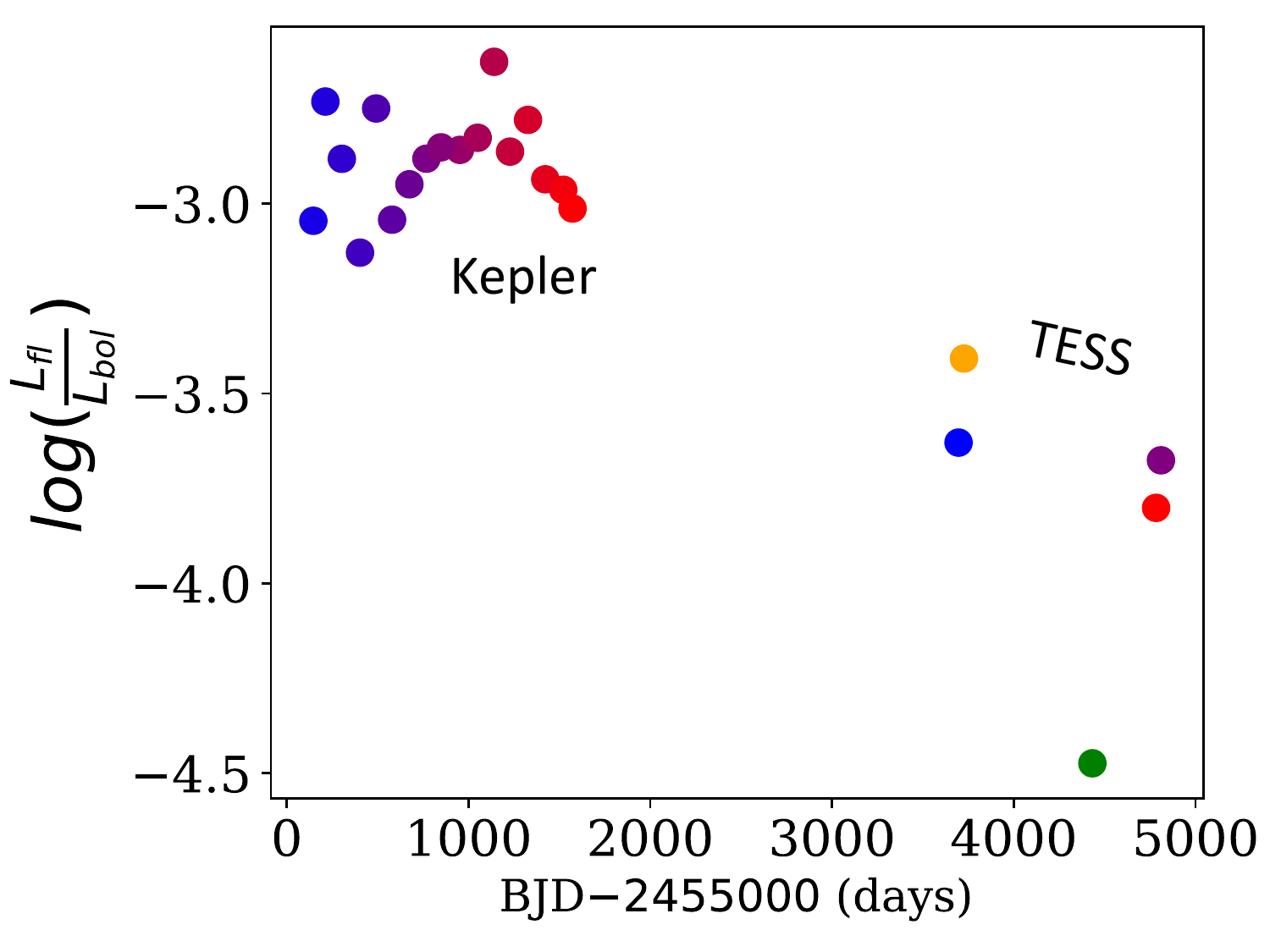}
\caption{Left: FFD intercepts $\beta$ (Eq.~\ref{eq:beta}) for both Kepler and TESS observations, with colors corresponding to Figure~\ref{fig:kepler_lc}, and Figure~\ref{fig:tessffd} for TESS. Error bars are derived from adding in quadrature the fitting error, and the statistical uncertainty derived in Section~\ref{sec:stats} (Figure~\ref{fig:ten_year}). Right: integrated flare power log($L_{fl}/L_{bol}$) for Kepler and TESS observations. There is a clear offset between the Kepler and the TESS data in both panels, which could be evidence for a long term trend in flare rate, but is also sensitive to the differences in cadence, S/N, and bandpass between Kepler and TESS.}
\label{fig:both_t_and_k}
\end{figure*}

While our re-analysis of the Kepler flare sample does not show the same long term trends as \citet{scoggins2019}, we do see an offset between the Kepler and TESS data. The median Kepler flare rate has a $\beta$ value nearly 0.5 dex higher than the TESS flare data in Figure \ref{fig:both_t_and_k}. Since $\beta$ is determined from the FFD, it should be robust to differences in the flare event energy sensitives between Kepler and TESS, as opposed to integrated flare power metrics such as the $L_{fl}/L_{bol}$ metric \citep[e.g.,][]{scoggins2019, feinstein_evolution_2024} in the right panel of Figure~\ref{fig:both_t_and_k}, which can also be impacted by the presence of single large energy flare events. While we prefer the $\beta$ metric for this reason, we show both for accurate comparison to literature studies and to demonstrate the trends present in the data are not metric dependent.
This offset in $\beta$ in Figure~\ref{fig:both_t_and_k} between Kepler and TESS could therefore be due to a long term trend in the flare rates, which takes place on decade timescales. 

However, we urge caution in the interpretation of the results in Figure \ref{fig:both_t_and_k} due to the competing effects of cadence and signal-to-noise. The Kepler data was sampled at 30-min cadence with a 0.95-m aperture \citep[e.g.,][]{koch2010}, while the TESS data was sampled at 2-min cadence with a 10-cm aperture \citep[][]{ricker_transiting_2015}. While faster sampling should yield better sensitivity to short duration flares \citep{howard2022}, the smaller aperture of TESS means that low amplitude (lower energy) flares are not detectable. As \citet{davenport2019} note, measuring flare activity via the FFD is preferable when comparing data with different detection thresholds, but this does not account for the impact of filter and flare temperature noted above.

\begin{figure}
    \centering
    \includegraphics[width=0.47\textwidth]{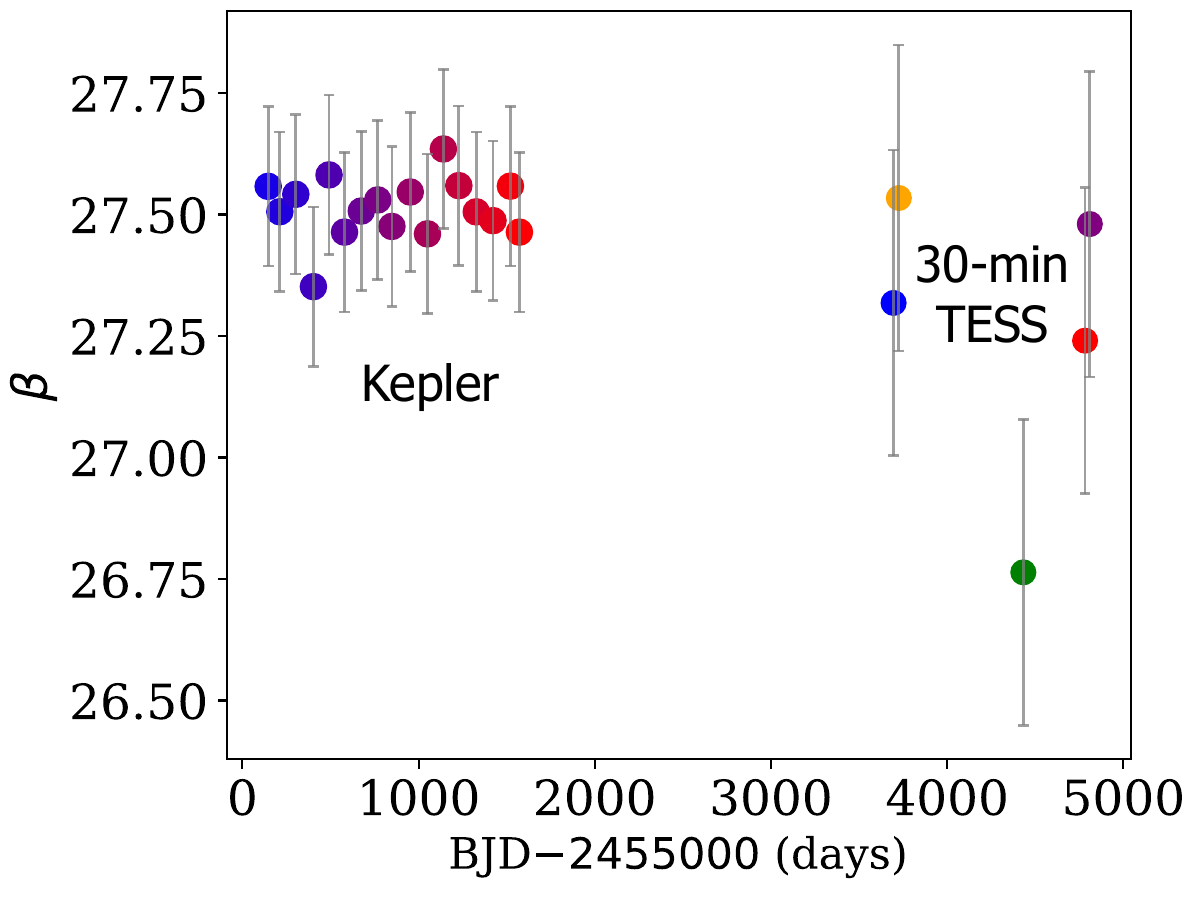}
    \caption{FFD intercepts $\beta$ (Eq.~\ref{eq:beta}) as in Figure~\ref{fig:both_t_and_k}, but with values for the TESS Sectors from the 2-min light curves binned down to 30-minute cadence, and flares identified using \texttt{FBEYE}. The offset between Kepler and TESS seen in Figure~\ref{fig:both_t_and_k} is present, though not as prevalent when considering similar cadence and identifications methods. However, differences in bandpasses still makes direct comparison difficult.}
    \label{fig:30min_TESS_intercepts}
\end{figure}

As an exercise to better match the TESS and Kepler datasets, we down sampled the TESS light curves from Section~\ref{sec:tess_data} to 30-minute cadence, and used \texttt{FBEYE} \citep{davenport2014b}, to replicate the flare identification used for Kepler data. These changes result in slightly different flare samples for the TESS data, with 174 flares identified by eye, versus 237 flares identified by the automated process. This difference is not unexpected, and is due to low energy flares being detected in the 2 min data. This impacts low energy end of the FFD, where we do not fit due to the sample incompleteness evident from our injection and recovery tests. We show the updated $\beta$ values over time for TESS in Figure~\ref{fig:30min_TESS_intercepts}.
We find that the $\beta$ value offset between the Kepler and TESS samples is smaller when using the 30-min light curves and manual flare identification. These results further demonstrate the necessity for consistency in flare sample methodologies when analyzing flare statistics over time, and especially between surveys.

Moreover, it is important to further consider the impact of the different bandpasses between Kepler and TESS, which affect our estimate of the flare event energies. \citet{davenport2020} demonstrated for GJ 1243 that the flares had energies within a few \% of each other from these two missions, assuming a typical 10,000 K blackbody flare spectrum. As in \citet{davenport2020}, we estimated the quiescent luminosity for TIC-272272592 in both the Kepler and TESS bands. As \citet{tovar-mendoza2022} demonstrates, flares look qualitatively identical between Kepler and TESS.
However, the \citet{davenport2020} comparison between Kepler and TESS assumed a simple 10,000 K blackbody spectrum for all flares, following the convention of many other studies \citep[e.g.][]{Notsu2013}. Recent work has shown that flare temperatures can deviate significantly from this 10,000 K assumption \citep[e.g.][]{kowalski2013}. This includes a clear change in the temperature over the profile of the flare (i.e., hotter temperatures at peak flare emission, with a gradual cooling profile), and reaching effective temperatures as high as 40,000 K for large flares \citep{howard2020}. 
\citet{raetz_long-term_2024} have similarly observed differing FFD slopes in comparisons between Kepler and TESS, though they note this was partially due to impacts from complex flare morphologies in their energy estimates. 
The correlation between flare energy and peak effective temperature, and the general profile of temperatures throughout flare events, is an active area of study at many wavelengths \citep[][Tovar Mendoza et al. in prep]{howard2023}. Both flare temperature profiles, and differences between the Kepler and TESS bandpasses, can contribute to the TESS sample having lower apparent flare activity.

In this work, we do not see convincing evidence of activity cycle driven, long term flare rate variation. For an active M3 dwarf like TIC-272272592, it is unclear what activity cycles should be expected. Current M dwarf dynamo models don't explore activity cycles for rapidly rotating stars, since activity cycles are typically associated with slower rotation \citep[e.g.,][]{bohm-vitense2007}. G and K stars with Solar-like dynamos have been shown to have activity cycles correlated with their rotational periods and ages \citep{olah_magnetic_2016}, with more rapidly rotating stars exhibiting shorter cycle durations. Extending such models would suggest an activity cycle of $< 1$yr for a star rotating at 1.2 days. For TIC-272272592, the Kepler data appears to rule out any strong activity cycle shorter than 2 years. This suggests that either the convective dynamo with such an active M dwarf either does not (or does not yet) generate activity cycles, or they operate on timescales outside the sensitivity of the data analyzed here. The convective dynamos of late type M dwarfs can show complex, multi-polar structure \citep{morin2010}, and variations manifested in activity indicators like flares may as a result be stochastic on shorter timescales.

Given the ongoing challenges in comparing flare activity between telescopes, particularly for samples that were not observed contemporaneously, long term trends are best constrained from a single telescope or survey. The TESS baseline alone is now approaching 7 years, and for stars in the continuous viewing zone, TESS will be able to provide the best glimpse into both the short term (Sector-to-Sector) and the long term (year-to-year) magnetic activity evolution via flares. We note this as the focus of our ongoing future work.

\section{Statistical Significance of Variations}\label{sec:stats}

As noted above, Sector 41 shows a clear deviation from the rest of the TESS Sectors in terms of both the number of large flares visible in the light curve, and the flare activity level recorded in the FFD $\beta$ value. This begs the question: how statistically significant is this low flare activity level, given the stochastic nature of flares, and observing a single TESS Sector of data? 

To determine this significance, we conduct a Monte Carlo style experiment with a theoretical power-law distribution of flares, randomly generated over a 10 year period.
For our 10 year baseline, we assume the null-hypothesis that there are no long term trends in the flare rate, and distribute the flares randomly throughout. We selected the parameters for the input power-law to match the observed flare rate for TIC-272272592, with a slope of $\alpha=-0.85$. We simulated a total of 7300 flares over the 10-year period, generating approximately 50 flares per 25-day window (i.e. close to the TESS flare rate). 
To avoid biasing our sample at low flare energies, we extended the power-law down to $10^{31.4} \, {\rm erg}$ events, though we only use flares above the mean $50\%$ completeness limit of the full sample of $\log$(Energy) = 32.05. 

\begin{figure*}[ht!]
    \centering
    \includegraphics[width=0.49\textwidth]{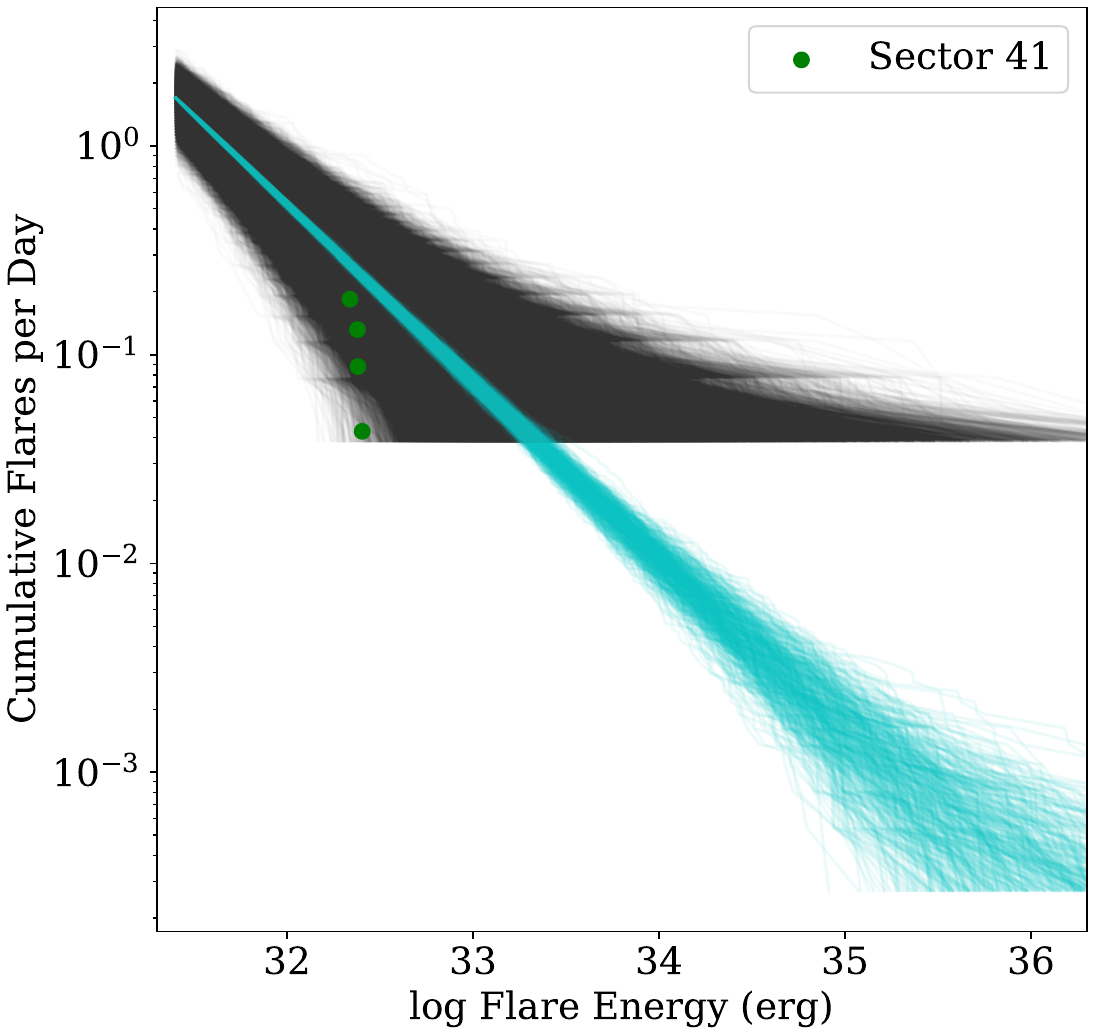}
    \includegraphics[width=0.49\textwidth]{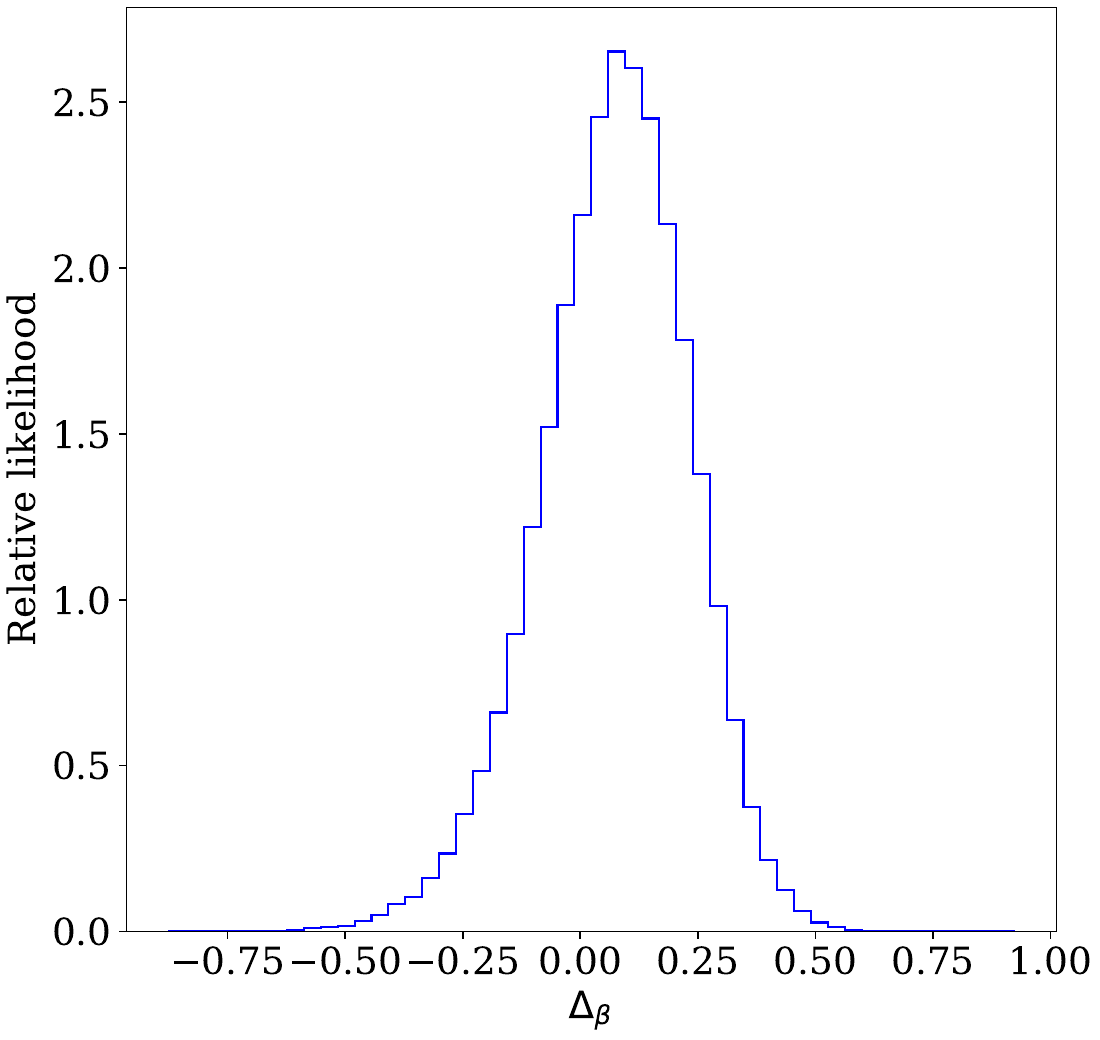}
    \caption{Left: results for drawing ten-year baseline power-law distributions and sub-sampling TESS sectors. The 500 individual power-law draws are shown in cyan. For each trial, 100 single TESS sector re-draws are shown in black. Shown in green is Sector 41, re-plotted from Figure~\ref{fig:tessffd} for direct comparison to the simulation}. Right: the distribution of the $\Delta_b$'s for each of the sector re-draws from the left panel.
    \label{fig:ten_year}
\end{figure*}

From this 10-year baseline, we generate mock TESS Sectors of 27 days, with a 1.12 day gap in the middle to simulate the telescope down-link period. We create 100 simulated TESS Sectors with random start times. We then re-generate the 10-year baseline 500 times, drawing 100 mock TESS Sectors from each realization. In total, we simulated 5000 years of flare data, generating 50,000 FFDs. The results of this Monte Carlo experiment are shown in Figure~\ref{fig:ten_year}, where the 500 simulated 10-year power-laws are shown in cyan, while the 50,000 Sector re-draws are shown in black. 

Due to the small variation in the 500 simulated 10-year samples due to Poisson number statistics, we fit each 10-year FFD independently. We then fit the 50,000 mock TESS Sector FFDs with our fixed-slope model, and we calculated the difference from the respective 10-year value, which we define as $\Delta_{\beta}$. This $\Delta_{\beta}$ quantifies the difference in the flare activity level determined from the FFD between the 10-year ``truth'' and the point-in-time estimate with a realistic TESS Sector.
The distribution of 50,000 $\Delta_{\beta}$'s from this experiment is shown in Figure~\ref{fig:ten_year}. We note that this distribution is not centered at $\Delta_{\beta}$=0, but rather at $\Delta_{\beta} = 0.08$, indicating that for a single TESS sector, the fit intercept $\beta$ is likely slightly {\it over} estimated, as the presence of even a single high energy flare will result in deviation of the FFD from a seemingly fixed power-law shape. A similar result was demonstrated in the single high outlier experiment for the star cluster mass function by \citet{wainer_panchromatic_2022}. 

In the left panel of Figure~\ref{fig:ten_year}, we overlay the four flares from TESS Sector 41 that are above our computed completeness limit (re-plotted from Figure~\ref{fig:tessffd}). From this comparison we can see that there are simulation draws which cover the parameter space occupied by Sector 41. While our simulations were designed to match the mean flare rate observed in TESS, and include no short- or long-term variation beyond random number statistics, the Sector 41 data clearly fall within the band of the simulated TESS Sectors. This indicates that the low flare rate observed in Sector 41 could plausibly be explained by the stochastic nature of flare events.
Quantitatively, the $\Delta_{\beta}$ distribution from our simulations in Figure~\ref{fig:ten_year} shows that the $\Delta_{\beta} = -0.51$ value observed in TESS Sector 41 is a $2.7\sigma$ outlier as compared to the other 4 Sectors. While this Sector 41 is low in flare activity, a single TESS Sector at this level is not enough to disprove the null hypothesis that the flare rate is constant over time. 

\begin{figure}
    \centering
    \includegraphics[width=0.47\textwidth]{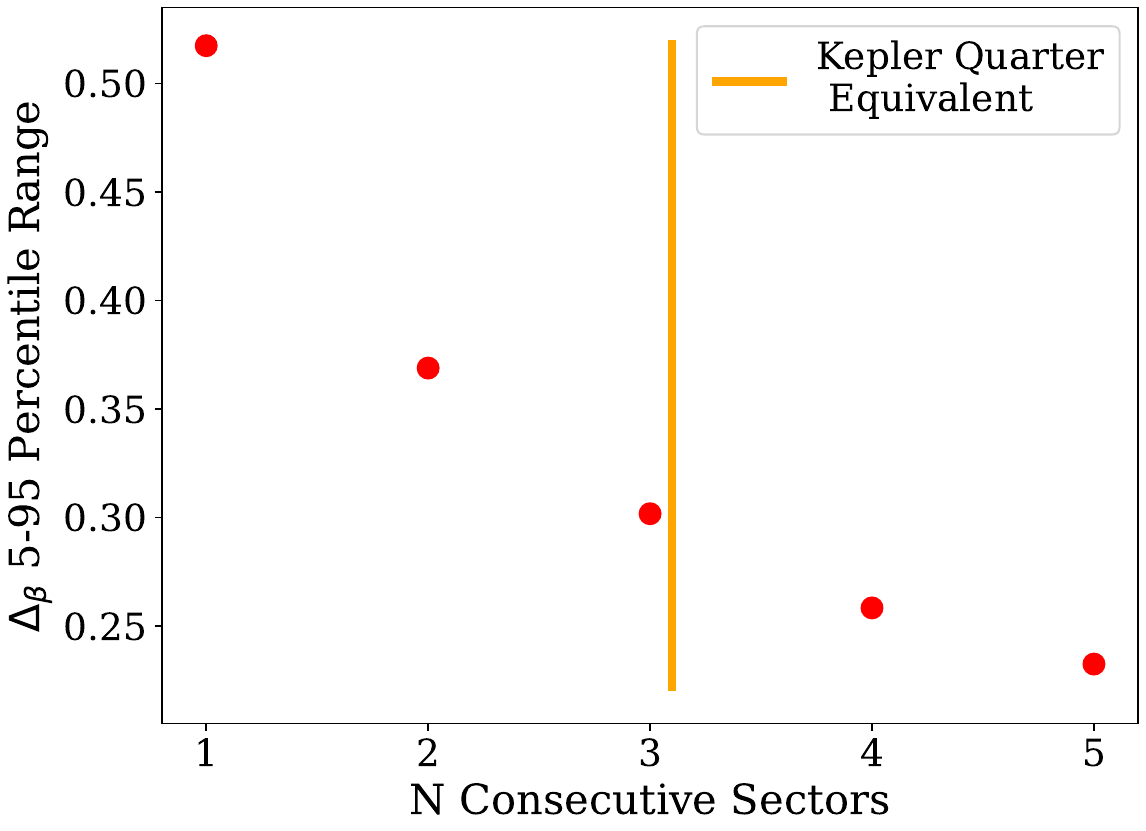}
    \caption{Width of the $\Delta_b$ distribution as a function of how many continuous sectors of data are sampled quantified by the 5th to 95th percentile range. The orange line represents how many consecutive TESS sectors are equivalent to one Kepler Quarter. We see an exponential decrease in width with the number of sectors, indicating that for stars with more sectors, better estimates of point in time flare rates can be measured.}
    \label{fig:width_dbs}
\end{figure}

Repeating this Monte Carlo exercise for the Kepler data, assuming each simulated Kepler Quarter was equivalent to 3 TESS Sectors, we find a 5th to 95th percentile range in the $\Delta_{\beta}$ distribution of 0.30. 
This spread is substantially smaller than for the TESS Sector simulation (5th to 95th percentile range of 0.52), and is in agreement with the uncertainty we expect from randomly drawing from a power law over a finite time period (Figure~\ref{fig:width_dbs}). Within the actual Kepler flare sample in Figure \ref{fig:both_t_and_k}, the maximum deviation from the median $\beta$ occurs in Quarter 4, with a $\Delta_{\beta}$ of 0.18. This is less than a $2\sigma$ deviation, using the tests presented here. For comparison, the 5th to 95th percentile range in $\beta$ values for the actual Kepler sample are 0.19, which is smaller than what we expect from the simulation.

To detect real variations in flare rate, we must understand the impact that the observing baseline has on our estimate of the FFD $\beta$. 
As shown above, longer baseline observations (e.g. a Kepler Quarter versus a TESS Sector) reduces the variation in the point-in-time $\beta$. From our ensemble of 10-year flare simulations, we can easily explore the impact of longer observing baselines. In  Figure~\ref{fig:width_dbs} we show the resulting 5th to 95th percentile range in the $\Delta_{\beta}$ distribution as a function of number of consecutive simulated TESS Sectors. As expected, we find that the more continuous Sectors of data used to determine the FFD, the more robustly the input power-law is recovered. The improvement in the $\Delta_{\beta}$ distribution in Figure~\ref{fig:width_dbs} follows a typical $\sqrt{N}$ behavior with increasing numbers of Sectors. This experiment highlights the value of studying flare activity evolution in the TESS Continuous Viewing Zone, where multiple TESS Sectors can be combined to improve the precision and therefore look for lower amplitude changes in flare activity over time. Conversely, using single TESS Sectors to constrain the point-in-time flare rate, we are only sensitive to larger amplitude changes in the activity level.

Lastly, we emphasize that the $\Delta_{\beta}$ precision demonstrated here is dependent on the specific flare rate of the simulated star, and that the true improvement in the FFD comes from increasing the numbers of flares in the sample. In other words, a star with a higher specific flare rate would have better $\Delta_{\beta}$ precision as a function of time. However, we choose to present this Monte Carlo result as a function of Sectors since this is a typical flaring M dwarf, and Sectors are the fundamental unit of time when using data from the TESS mission.

\section{Conclusion}
\label{sec:end}


We have demonstrated that flares provide a novel method for exploring solar and stellar activity cycles. Over a decade timescale, X-ray flares from the Sun change their occurrence rate by more than an order of magnitude between activity maximum and minimum. 
As flares are relatively easy to observe and quantify compared with other activity cycle tracers or surface magnetism features (e.g. starspots), they are a uniquely capable metric for constraining activity cycles in large surveys. 

Following one of the first candidate flare rate variations from \citet{scoggins2019}, we have presented a reanalysis of the Kepler light curves, and one of the first flare studies from new TESS light curves, for the rapidly rotating M dwarf, TIC-272272592. The slow decline in flare rate previously claimed is not found in our data, and careful analysis of the FFD shows that the Kepler flare data is remarkably stable over 4 years. We see a 0.5 dex difference between the $\beta$ values of the Kepler sample compared to TESS, which we show to be dependant on both telescope and light curve cadence differences. Considering the TESS flare rates measured using 2-min versus 30-min light curves in \S 4.2, we emphasize the difficulties in comparing flare samples between surveys and with differing methodologies.  

Within the TESS sample, we find a $2.7\sigma$ detection of dramatic short timescale (month-to-month) flare rate variation for TIC-272272592, which could be the result of sampling statistics for a power-law Poisson process within a given observation window. Our Monte Carlo tests confirm that detecting small amplitude variations in flare rate requires more than a single TESS Sector to constrain the point-in-time flare activity level for typical active M dwarfs. High amplitude changes in flare rate are detectable using single Sectors. The TESS Continuous Viewing Zone is a promising region to conduct more sensitive studies for flare rate variations, and we are currently working on this for a follow-up project. 

The methods developed here are the first in our exploration of long term flare behavior from missions like Kepler, TESS, and beyond. Constraining both the short term and long term variations in flare rate is important for understanding activity cycle behavior, particularly for young stars, and for predicting the impact that flares have on e.g. exoplanet transit detection efficiency.
Wide field exoplanet surveys like TESS allow ensemble flare studies for thousands of nearby stars on decade timescales, which will continue to shed light on the constancy of their variability over time.


\vspace{1.75cm}

JRAD acknowledges support from the DiRAC Institute in the Department of Astronomy at the University of Washington. The DIRAC Institute is supported through generous gifts from the Charles and Lisa Simonyi Fund for Arts and Sciences, and the Washington Research Foundation. ADF acknowledges support by NASA through the NASA Hubble Fellowship grant HST-HF2-51530 awarded by STScI.

This research was supported by the National Aeronautics and Space Administration (NASA) under grant number 80NSSC21K0362 from the TESS Cycle 3 Guest Investigator Program, grant number 80NSSC23K0155 from the TESS Cycle 5 Guest Investigator Program, and award number 80NSSC23K0970 from the Astrophysics Data Analysis Program (ADAP). TW acknowledges support from NASA ATP grant 80NSSC24K0768. This research has made use of NASA's Astrophysics Data System.


\software{\texttt{astropy} \citep{astropy:2013, astropy:2018, astropy:2022}, \texttt{Jupyter} \citep{2007CSE.....9c..21P, kluyver2016jupyter}, \texttt{matplotlib} \citep{Hunter:2007}, \texttt{numpy} \citep{numpy}, \texttt{pandas} \citep{mckinney-proc-scipy-2010, pandas_10697587}, \texttt{python} \citep{python}, \texttt{scipy} \citep{2020SciPy-NMeth, scipy_12522488}, \texttt{astroquery} \citep{2019AJ....157...98G, astroquery_10799414}, \texttt{Cython} \citep{cython:2011}, \texttt{eleanor} \citep{Feinstein+2019:2019PASP..131i4502F}, \texttt{emcee} \citep{emcee-Foreman-Mackey-2013, emcee_10996751}, \texttt{h5py} \citep{collette_python_hdf5_2014, h5py_7560547}, \texttt{Lightkurve} \citep{2018ascl.soft12013L, Lightkurve_4558241}, \texttt{scikit-image} \citep{scikit-image}, \texttt{scikit-learn} \citep{scikit-learn, sklearn_api, scikit-learn_10666857}, \texttt{tqdm} \citep{tqdm_8233425} and \texttt{Photutils} \citep{Photutils_12585239}.
Software citation information aggregated using \texttt{\href{https://www.tomwagg.com/software-citation-station/}{The Software Citation Station}} \citep{software-citation-station-paper, software-citation-station-zenodo}.}

\bibliography{references.bib, tobins_references.bib, refs_additional.bib, software.bib}

\begin{thebibliography}{}
\expandafter\ifx\csname natexlab\endcsname\relax\def\natexlab#1{#1}\fi
\providecommand{\url}[1]{\href{#1}{#1}}
\providecommand{\dodoi}[1]{doi:~\href{http://doi.org/#1}{\nolinkurl{#1}}}
\providecommand{\doeprint}[1]{\href{http://ascl.net/#1}{\nolinkurl{http://ascl.net/#1}}}
\providecommand{\doarXiv}[1]{\href{https://arxiv.org/abs/#1}{\nolinkurl{https://arxiv.org/abs/#1}}}

\bibitem[{{Aigrain} {et~al.}(2015){Aigrain}, {Llama}, {Ceillier}, {Chagas}, {Davenport}, {Garc{\'{\i}}a}, {Hay}, {Lanza}, {McQuillan}, {Mazeh}, {de Medeiros}, {Nielsen}, \& {Reinhold}}]{aigrain2015}
{Aigrain}, S., {Llama}, J., {Ceillier}, T., {et~al.} 2015, \mnras, 450, 3211, \dodoi{10.1093/mnras/stv853}

\bibitem[{{Astropy Collaboration} {et~al.}(2013){Astropy Collaboration}, {Robitaille}, {Tollerud}, {Greenfield}, {Droettboom}, {Bray}, {Aldcroft}, {Davis}, {Ginsburg}, {Price-Whelan}, {Kerzendorf}, {Conley}, {Crighton}, {Barbary}, {Muna}, {Ferguson}, {Grollier}, {Parikh}, {Nair}, {Unther}, {Deil}, {Woillez}, {Conseil}, {Kramer}, {Turner}, {Singer}, {Fox}, {Weaver}, {Zabalza}, {Edwards}, {Azalee Bostroem}, {Burke}, {Casey}, {Crawford}, {Dencheva}, {Ely}, {Jenness}, {Labrie}, {Lim}, {Pierfederici}, {Pontzen}, {Ptak}, {Refsdal}, {Servillat}, \& {Streicher}}]{astropy:2013}
{Astropy Collaboration}, {Robitaille}, T.~P., {Tollerud}, E.~J., {et~al.} 2013, \aap, 558, A33, \dodoi{10.1051/0004-6361/201322068}

\bibitem[{{Astropy Collaboration} {et~al.}(2018){Astropy Collaboration}, {Price-Whelan}, {Sip{\H{o}}cz}, {G{\"u}nther}, {Lim}, {Crawford}, {Conseil}, {Shupe}, {Craig}, {Dencheva}, {Ginsburg}, {Vand erPlas}, {Bradley}, {P{\'e}rez-Su{\'a}rez}, {de Val-Borro}, {Aldcroft}, {Cruz}, {Robitaille}, {Tollerud}, {Ardelean}, {Babej}, {Bach}, {Bachetti}, {Bakanov}, {Bamford}, {Barentsen}, {Barmby}, {Baumbach}, {Berry}, {Biscani}, {Boquien}, {Bostroem}, {Bouma}, {Brammer}, {Bray}, {Breytenbach}, {Buddelmeijer}, {Burke}, {Calderone}, {Cano Rodr{\'\i}guez}, {Cara}, {Cardoso}, {Cheedella}, {Copin}, {Corrales}, {Crichton}, {D'Avella}, {Deil}, {Depagne}, {Dietrich}, {Donath}, {Droettboom}, {Earl}, {Erben}, {Fabbro}, {Ferreira}, {Finethy}, {Fox}, {Garrison}, {Gibbons}, {Goldstein}, {Gommers}, {Greco}, {Greenfield}, {Groener}, {Grollier}, {Hagen}, {Hirst}, {Homeier}, {Horton}, {Hosseinzadeh}, {Hu}, {Hunkeler}, {Ivezi{\'c}}, {Jain}, {Jenness}, {Kanarek}, {Kendrew}, {Kern}, {Kerzendorf}, {Khvalko}, {King}, {Kirkby}, {Kulkarni},
  {Kumar}, {Lee}, {Lenz}, {Littlefair}, {Ma}, {Macleod}, {Mastropietro}, {McCully}, {Montagnac}, {Morris}, {Mueller}, {Mumford}, {Muna}, {Murphy}, {Nelson}, {Nguyen}, {Ninan}, {N{\"o}the}, {Ogaz}, {Oh}, {Parejko}, {Parley}, {Pascual}, {Patil}, {Patil}, {Plunkett}, {Prochaska}, {Rastogi}, {Reddy Janga}, {Sabater}, {Sakurikar}, {Seifert}, {Sherbert}, {Sherwood-Taylor}, {Shih}, {Sick}, {Silbiger}, {Singanamalla}, {Singer}, {Sladen}, {Sooley}, {Sornarajah}, {Streicher}, {Teuben}, {Thomas}, {Tremblay}, {Turner}, {Terr{\'o}n}, {van Kerkwijk}, {de la Vega}, {Watkins}, {Weaver}, {Whitmore}, {Woillez}, {Zabalza}, \& {Astropy Contributors}}]{astropy:2018}
{Astropy Collaboration}, {Price-Whelan}, A.~M., {Sip{\H{o}}cz}, B.~M., {et~al.} 2018, \aj, 156, 123, \dodoi{10.3847/1538-3881/aabc4f}

\bibitem[{{Astropy Collaboration} {et~al.}(2022){Astropy Collaboration}, {Price-Whelan}, {Lim}, {Earl}, {Starkman}, {Bradley}, {Shupe}, {Patil}, {Corrales}, {Brasseur}, {N{"o}the}, {Donath}, {Tollerud}, {Morris}, {Ginsburg}, {Vaher}, {Weaver}, {Tocknell}, {Jamieson}, {van Kerkwijk}, {Robitaille}, {Merry}, {Bachetti}, {G{"u}nther}, {Aldcroft}, {Alvarado-Montes}, {Archibald}, {B{'o}di}, {Bapat}, {Barentsen}, {Baz{'a}n}, {Biswas}, {Boquien}, {Burke}, {Cara}, {Cara}, {Conroy}, {Conseil}, {Craig}, {Cross}, {Cruz}, {D'Eugenio}, {Dencheva}, {Devillepoix}, {Dietrich}, {Eigenbrot}, {Erben}, {Ferreira}, {Foreman-Mackey}, {Fox}, {Freij}, {Garg}, {Geda}, {Glattly}, {Gondhalekar}, {Gordon}, {Grant}, {Greenfield}, {Groener}, {Guest}, {Gurovich}, {Handberg}, {Hart}, {Hatfield-Dodds}, {Homeier}, {Hosseinzadeh}, {Jenness}, {Jones}, {Joseph}, {Kalmbach}, {Karamehmetoglu}, {Ka{l}uszy{'n}ski}, {Kelley}, {Kern}, {Kerzendorf}, {Koch}, {Kulumani}, {Lee}, {Ly}, {Ma}, {MacBride}, {Maljaars}, {Muna}, {Murphy}, {Norman}, {O'Steen},
  {Oman}, {Pacifici}, {Pascual}, {Pascual-Granado}, {Patil}, {Perren}, {Pickering}, {Rastogi}, {Roulston}, {Ryan}, {Rykoff}, {Sabater}, {Sakurikar}, {Salgado}, {Sanghi}, {Saunders}, {Savchenko}, {Schwardt}, {Seifert-Eckert}, {Shih}, {Jain}, {Shukla}, {Sick}, {Simpson}, {Singanamalla}, {Singer}, {Singhal}, {Sinha}, {Sip{H{o}}cz}, {Spitler}, {Stansby}, {Streicher}, {{{S}}umak}, {Swinbank}, {Taranu}, {Tewary}, {Tremblay}, {Val-Borro}, {Van Kooten}, {Vasovi{'c}}, {Verma}, {de Miranda Cardoso}, {Williams}, {Wilson}, {Winkel}, {Wood-Vasey}, {Xue}, {Yoachim}, {Zhang}, {Zonca}, \& {Astropy Project Contributors}}]{astropy:2022}
{Astropy Collaboration}, {Price-Whelan}, A.~M., {Lim}, P.~L., {et~al.} 2022, apj, 935, 167, \dodoi{10.3847/1538-4357/ac7c74}

\bibitem[{{Babcock}(1961)}]{babcock1961}
{Babcock}, H.~W. 1961, \apj, 133, 572, \dodoi{10.1086/147060}

\bibitem[{{Baliunas} {et~al.}(1995){Baliunas}, {Donahue}, {Soon}, {Horne}, {Frazer}, {Woodard-Eklund}, {Bradford}, {Rao}, {Wilson}, {Zhang}, {Bennett}, {Briggs}, {Carroll}, {Duncan}, {Figueroa}, {Lanning}, {Misch}, {Mueller}, {Noyes}, {Poppe}, {Porter}, {Robinson}, {Russell}, {Shelton}, {Soyumer}, {Vaughan}, \& {Whitney}}]{baliunas1995}
{Baliunas}, S.~L., {Donahue}, R.~A., {Soon}, W.~H., {et~al.} 1995, \apj, 438, 269, \dodoi{10.1086/175072}

\bibitem[{Barentsen {et~al.}(2021)Barentsen, Hedges, Vinícius, Saunders, gully, Lee, rebekah9969, Bell, Sagear, Barclay, Burke, KenMighell, jcsmithhere, III, Cody, Foreman-Mackey, Turtelboom, Davies, Colman, Zhang, Sundaram, Vincello, Daniel, Homeier, Higgins00, Bado, Martínez-Palomera, Cui, Carvalho, \& Hall}]{Lightkurve_4558241}
Barentsen, G., Hedges, C., Vinícius, Z., {et~al.} 2021, lightkurve/lightkurve: Lightkurve v2.0.3, v2.0.3,  Zenodo, \dodoi{10.5281/zenodo.4558241}

\bibitem[{{Baum} {et~al.}(2022){Baum}, {Wright}, {Luhn}, \& {Isaacson}}]{baum2022}
{Baum}, A.~C., {Wright}, J.~T., {Luhn}, J.~K., \& {Isaacson}, H. 2022, \aj, 163, 183, \dodoi{10.3847/1538-3881/ac5683}

\bibitem[{Behnel {et~al.}(2011)Behnel, Bradshaw, Citro, Dalcin, Seljebotn, \& Smith}]{cython:2011}
Behnel, S., Bradshaw, R., Citro, C., {et~al.} 2011, Computing in Science Engineering, 13, 31, \dodoi{10.1109/MCSE.2010.118}

\bibitem[{{B{\"o}hm-Vitense}(2007)}]{bohm-vitense2007}
{B{\"o}hm-Vitense}, E. 2007, \apj, 657, 486, \dodoi{10.1086/510482}

\bibitem[{{Borucki} {et~al.}(2010){Borucki}, {Koch}, {Basri}, {Batalha}, {Brown}, {Caldwell}, {Caldwell}, {Christensen-Dalsgaard}, {Cochran}, {DeVore}, {Dunham}, {Dupree}, {Gautier}, {Geary}, {Gilliland}, {Gould}, {Howell}, {Jenkins}, {Kondo}, {Latham}, {Marcy}, {Meibom}, {Kjeldsen}, {Lissauer}, {Monet}, {Morrison}, {Sasselov}, {Tarter}, {Boss}, {Brownlee}, {Owen}, {Buzasi}, {Charbonneau}, {Doyle}, {Fortney}, {Ford}, {Holman}, {Seager}, {Steffen}, {Welsh}, {Rowe}, {Anderson}, {Buchhave}, {Ciardi}, {Walkowicz}, {Sherry}, {Horch}, {Isaacson}, {Everett}, {Fischer}, {Torres}, {Johnson}, {Endl}, {MacQueen}, {Bryson}, {Dotson}, {Haas}, {Kolodziejczak}, {Van Cleve}, {Chandrasekaran}, {Twicken}, {Quintana}, {Clarke}, {Allen}, {Li}, {Wu}, {Tenenbaum}, {Verner}, {Bruhweiler}, {Barnes}, \& {Prsa}}]{borucki2010}
{Borucki}, W.~J., {Koch}, D., {Basri}, G., {et~al.} 2010, Science, 327, 977, \dodoi{10.1126/science.1185402}

\bibitem[{Bradley {et~al.}(2024)Bradley, Sipőcz, Robitaille, Tollerud, Vinícius, Deil, Barbary, Wilson, Busko, Donath, Günther, Cara, Lim, Meßlinger, Burnett, Conseil, Droettboom, Bostroem, Bray, Bratholm, Jamieson, Ginsburg, Barentsen, Craig, Pascual, Rathi, Perrin, Morris, \& Perren}]{Photutils_12585239}
Bradley, L., Sipőcz, B., Robitaille, T., {et~al.} 2024, astropy/photutils: 1.13.0, 1.13.0,  Zenodo, \dodoi{10.5281/zenodo.12585239}

\bibitem[{Bressan {et~al.}(2012)Bressan, Marigo, Girardi, Salasnich, Dal~Cero, Rubele, \& Nanni}]{bressan_parsec_2012}
Bressan, A., Marigo, P., Girardi, L., {et~al.} 2012, Monthly Notices of the Royal Astronomical Society, 427, 127, \dodoi{10.1111/j.1365-2966.2012.21948.x}

\bibitem[{Buitinck {et~al.}(2013)Buitinck, Louppe, Blondel, Pedregosa, Mueller, Grisel, Niculae, Prettenhofer, Gramfort, Grobler, Layton, VanderPlas, Joly, Holt, \& Varoquaux}]{sklearn_api}
Buitinck, L., Louppe, G., Blondel, M., {et~al.} 2013, in ECML PKDD Workshop: Languages for Data Mining and Machine Learning, 108--122

\bibitem[{Collette(2013)}]{collette_python_hdf5_2014}
Collette, A. 2013, Python and HDF5 (O'Reilly)

\bibitem[{Collette {et~al.}(2023)Collette, Kluyver, Caswell, Tocknell, Kieffer, Jelenak, Scopatz, Dale, Chen, VINCENT, Einhorn, payno, juliagarriga, Sciarelli, Valls, Ghosh, Pedersen, Kittisopikul, jakirkham, Raspaud, Danilevski, Abbasi, Readey, Mühlbauer, Paramonov, Chan, Schepper, Solé, jialin, \& Guest}]{h5py_7560547}
Collette, A., Kluyver, T., Caswell, T.~A., {et~al.} 2023, h5py/h5py: 3.8.0, 3.8.0,  Zenodo, \dodoi{10.5281/zenodo.7560547}

\bibitem[{da~Costa-Luis {et~al.}(2023)da~Costa-Luis, Larroque, Altendorf, Mary, richardsheridan, Korobov, Yorav-Raphael, Ivanov, Bargull, Rodrigues, Chen, Lee, Newey, CrazyPython, JC, Zugnoni, Pagel, mjstevens777, Dektyarev, Rothberg, Plavin, Dill, FichteFoll, Sturm, HeoHeo, van Kemenade, McCracken, MapleCCC, Nordlund, \& Boyle}]{tqdm_8233425}
da~Costa-Luis, C., Larroque, S.~K., Altendorf, K., {et~al.} 2023, {tqdm: A fast, Extensible Progress Bar for Python and CLI}, v4.66.1,  Zenodo, \dodoi{10.5281/zenodo.8233425}

\bibitem[{{Davenport}(2016)}]{davenport2016}
{Davenport}, J.~R.~A. 2016, \apj, 829, 23, \dodoi{10.3847/0004-637X/829/1/23}

\bibitem[{{Davenport} {et~al.}(2019){Davenport}, {Covey}, {Clarke}, {Boeck}, {Cornet}, \& {Hawley}}]{davenport2019}
{Davenport}, J. R.~A., {Covey}, K.~R., {Clarke}, R.~W., {et~al.} 2019, \apj, 871, 241, \dodoi{10.3847/1538-4357/aafb76}

\bibitem[{{Davenport} {et~al.}(2020){Davenport}, {Mendoza}, \& {Hawley}}]{davenport2020}
{Davenport}, J. R.~A., {Mendoza}, G.~T., \& {Hawley}, S.~L. 2020, \aj, 160, 36, \dodoi{10.3847/1538-3881/ab9536}

\bibitem[{{Davenport} {et~al.}(2014){Davenport}, {Hawley}, {Hebb}, {Wisniewski}, {Kowalski}, {Johnson}, {Malatesta}, {Peraza}, {Keil}, {Silverberg}, {Jansen}, {Scheffler}, {Berdis}, {Larsen}, \& {Hilton}}]{davenport2014b}
{Davenport}, J.~R.~A., {Hawley}, S.~L., {Hebb}, L., {et~al.} 2014, \apj, 797, 122, \dodoi{10.1088/0004-637X/797/2/122}

\bibitem[{Duncan {et~al.}(1991)Duncan, Vaughan, Wilson, Preston, Frazer, Lanning, Misch, Mueller, Soyumer, Woodard, Baliunas, Noyes, Hartmann, Porter, Zwaan, Middelkoop, Rutten, \& Mihalas}]{duncan_ca_1991}
Duncan, D.~K., Vaughan, A.~H., Wilson, O.~C., {et~al.} 1991, The Astrophysical Journal Supplement Series, 76, 383, \dodoi{10.1086/191572}

\bibitem[{{Eddy}(1980)}]{eddy1980}
{Eddy}, J.~A. 1980, in The Ancient Sun: Fossil Record in the Earth, Moon and Meteorites, ed. R.~O. {Pepin}, J.~A. {Eddy}, \& R.~B. {Merrill}, 119--134

\bibitem[{{Egeland} {et~al.}(2017){Egeland}, {Soon}, {Baliunas}, {Hall}, {Pevtsov}, \& {Bertello}}]{egeland2017}
{Egeland}, R., {Soon}, W., {Baliunas}, S., {et~al.} 2017, \apj, 835, 25, \dodoi{10.3847/1538-4357/835/1/25}

\bibitem[{Feinstein {et~al.}(2020{\natexlab{a}})Feinstein, Montet, \& Ansdell}]{feinstein_stella_2020}
Feinstein, A., Montet, B., \& Ansdell, M. 2020{\natexlab{a}}, The Journal of Open Source Software, 5, 2347, \dodoi{10.21105/joss.02347}

\bibitem[{Feinstein {et~al.}(2020{\natexlab{b}})Feinstein, Montet, Ansdell, Nord, Bean, Günther, Gully-Santiago, \& Schlieder}]{feinstein_flare_2020}
Feinstein, A.~D., Montet, B.~T., Ansdell, M., {et~al.} 2020{\natexlab{b}}, The Astronomical Journal, 160, 219, \dodoi{10.3847/1538-3881/abac0a}

\bibitem[{Feinstein {et~al.}(2024)Feinstein, Seligman, France, Gagné, \& Kowalski}]{feinstein_evolution_2024}
Feinstein, A.~D., Seligman, D.~Z., France, K., Gagné, J., \& Kowalski, A. 2024, Evolution of {Flare} {Activity} in {GKM} {Stars} {Younger} than 300 {Myr} over {Five} {Years} of {TESS} {Observations}, \dodoi{10.48550/arXiv.2405.00850}

\bibitem[{Feinstein {et~al.}(2022)Feinstein, Seligman, Günther, \& Adams}]{feinstein_testing_2022}
Feinstein, A.~D., Seligman, D.~Z., Günther, M.~N., \& Adams, F.~C. 2022, The Astrophysical Journal, 925, L9, \dodoi{10.3847/2041-8213/ac4b5e}

\bibitem[{{Feinstein} {et~al.}(2019){Feinstein}, {Montet}, {Foreman-Mackey}, {Bedell}, {Saunders}, {Bean}, {Christiansen}, {Hedges}, {Luger}, {Scolnic}, \& {Cardoso}}]{Feinstein+2019:2019PASP..131i4502F}
{Feinstein}, A.~D., {Montet}, B.~T., {Foreman-Mackey}, D., {et~al.} 2019, \pasp, 131, 094502, \dodoi{10.1088/1538-3873/ab291c}

\bibitem[{Foreman-Mackey(2018)}]{foreman-mackey_scalable_2018}
Foreman-Mackey, D. 2018, Research Notes of the American Astronomical Society, 2, 31, \dodoi{10.3847/2515-5172/aaaf6c}

\bibitem[{{Foreman-Mackey} {et~al.}(2013){Foreman-Mackey}, {Hogg}, {Lang}, \& {Goodman}}]{emcee-Foreman-Mackey-2013}
{Foreman-Mackey}, D., {Hogg}, D.~W., {Lang}, D., \& {Goodman}, J. 2013, \pasp, 125, 306, \dodoi{10.1086/670067}

\bibitem[{Foreman-Mackey {et~al.}(2024)Foreman-Mackey, Farr, Archibald, Tollerud, Hogg, Nelson, Kern, Sanders, Williams, Lang, Sinha, Béky, joezuntz, Price-Whelan, Dill, de~Val-Borro, Vousden, Ilya, Abril-Pla, Tazzari, Bouvard, Erhardt, Walker, Watkins, Martin, Bradshaw, Mohammed, \& Hoffimann}]{emcee_10996751}
Foreman-Mackey, D., Farr, W.~M., Archibald, A., {et~al.} 2024, dfm/emcee: v3.1.6, v3.1.6,  Zenodo, \dodoi{10.5281/zenodo.10996751}

\bibitem[{Foukal \& Lean(1988)}]{foukal_magnetic_1988}
Foukal, P., \& Lean, J. 1988, The Astrophysical Journal, 328, 347, \dodoi{10.1086/166297}

\bibitem[{Gao {et~al.}(2022)Gao, Liu, Yang, \& Zhou}]{gao_correcting_2022}
Gao, D.-Y., Liu, H.-G., Yang, M., \& Zhou, J.-L. 2022, The Astronomical Journal, 164, 213, \dodoi{10.3847/1538-3881/ac937e}

\bibitem[{{Ginsburg} {et~al.}(2019){Ginsburg}, {Sip{\H o}cz}, {Brasseur}, {Cowperthwaite}, {Craig}, {Deil}, {Guillochon}, {Guzman}, {Liedtke}, {Lian Lim}, {Lockhart}, {Mommert}, {Morris}, {Norman}, {Parikh}, {Persson}, {Robitaille}, {Segovia}, {Singer}, {Tollerud}, {de Val-Borro}, {Valtchanov}, {Woillez}, {The Astroquery collaboration}, \& {a subset of the astropy collaboration}}]{2019AJ....157...98G}
{Ginsburg}, A., {Sip{\H o}cz}, B.~M., {Brasseur}, C.~E., {et~al.} 2019, \aj, 157, 98, \dodoi{10.3847/1538-3881/aafc33}

\bibitem[{Ginsburg {et~al.}(2024)Ginsburg, Sipőcz, Brasseur, Parikh, jcsegovia, Groener, Norman, derdon, Kelley, Robitaille, Lim, Vaher, Deil, Mommert, Medina, Tollerud, Nilsson, Baumann, Craig, de~Val-Borro, Weaver, jespinosaar, Davies, Adeleke, Cowboy, Persson, Dempsey, syed gilani, Mesh, \& Mirocha}]{astroquery_10799414}
Ginsburg, A., Sipőcz, B., Brasseur, C.~E., {et~al.} 2024, astropy/astroquery: v0.4.7, v0.4.7,  Zenodo, \dodoi{10.5281/zenodo.10799414}

\bibitem[{Gommers {et~al.}(2024)Gommers, Virtanen, Haberland, Burovski, Weckesser, Reddy, Oliphant, Cournapeau, Nelson, alexbrc, Roy, Peterson, Polat, Wilson, endolith, Mayorov, van~der Walt, Brett, Laxalde, Larson, Sakai, Millman, Lars, peterbell10, Carey, van Mulbregt, Colley, Bowhay, eric jones, \& Striega}]{scipy_12522488}
Gommers, R., Virtanen, P., Haberland, M., {et~al.} 2024, scipy/scipy: SciPy 1.14.0, v1.14.0,  Zenodo, \dodoi{10.5281/zenodo.12522488}

\bibitem[{Grisel {et~al.}(2024)Grisel, Mueller, Lars, Gramfort, Louppe, Fan, Prettenhofer, Blondel, Niculae, Nothman, Joly, Lemaitre, Estève, Vanderplas, du~Boisberranger, manoj kumar, Qin, Hug, Varoquaux, Layton, Jalali, (Venkat)~Raghav, Schönberger, Yurchak, Jerphanion, Liu, la~Tour, Lorentzen, Li, \& Marmo}]{scikit-learn_10666857}
Grisel, O., Mueller, A., Lars, {et~al.} 2024, {scikit-learn/scikit-learn: Scikit-learn 1.4.1.post1}, 1.4.1.post1,  Zenodo, \dodoi{10.5281/zenodo.10666857}

\bibitem[{{G{\"u}nther} {et~al.}(2020){G{\"u}nther}, {Zhan}, {Seager}, {Rimmer}, {Ranjan}, {Stassun}, {Oelkers}, {Daylan}, {Newton}, {Kristiansen}, {Olah}, {Gillen}, {Rappaport}, {Ricker}, {Vanderspek}, {Latham}, {Winn}, {Jenkins}, {Glidden}, {Fausnaugh}, {Levine}, {Dittmann}, {Quinn}, {Krishnamurthy}, \& {Ting}}]{guenther20}
{G{\"u}nther}, M.~N., {Zhan}, Z., {Seager}, S., {et~al.} 2020, \aj, 159, 60, \dodoi{10.3847/1538-3881/ab5d3a}

\bibitem[{Hall {et~al.}(2007)Hall, Lockwood, \& Skiff}]{hall_activity_2007}
Hall, J.~C., Lockwood, G.~W., \& Skiff, B.~A. 2007, The Astronomical Journal, 133, 862, \dodoi{10.1086/510356}

\bibitem[{{Hawley} {et~al.}(2014){Hawley}, {Davenport}, {Kowalski}, {Wisniewski}, {Hebb}, {Deitrick}, \& {Hilton}}]{hawley2014}
{Hawley}, S.~L., {Davenport}, J.~R.~A., {Kowalski}, A.~F., {et~al.} 2014, \apj, 797, 121, \dodoi{10.1088/0004-637X/797/2/121}

\bibitem[{{Hilton} {et~al.}(2011){Hilton}, {Hawley}, {Kowalski}, \& {Holtzman}}]{hilton2011}
{Hilton}, E.~J., {Hawley}, S.~L., {Kowalski}, A.~F., \& {Holtzman}, J. 2011, in Astronomical Society of the Pacific Conference Series, Vol. 448, 16th Cambridge Workshop on Cool Stars, Stellar Systems, and the Sun, ed. C.~{Johns-Krull}, M.~K. {Browning}, \& A.~A. {West}, 197

\bibitem[{{Howard} {et~al.}(2019){Howard}, {Corbett}, {Law}, {Ratzloff}, {Glazier}, {Fors}, {del Ser}, \& {Haislip}}]{howard2019}
{Howard}, W.~S., {Corbett}, H., {Law}, N.~M., {et~al.} 2019, \apj, 881, 9, \dodoi{10.3847/1538-4357/ab2767}

\bibitem[{{Howard} \& {MacGregor}(2022)}]{howard2022}
{Howard}, W.~S., \& {MacGregor}, M.~A. 2022, \apj, 926, 204, \dodoi{10.3847/1538-4357/ac426e}

\bibitem[{{Howard} {et~al.}(2020){Howard}, {Corbett}, {Law}, {Ratzloff}, {Galliher}, {Glazier}, {Gonzalez}, {Vasquez Soto}, {Fors}, {del Ser}, \& {Haislip}}]{howard2020}
{Howard}, W.~S., {Corbett}, H., {Law}, N.~M., {et~al.} 2020, \apj, 902, 115, \dodoi{10.3847/1538-4357/abb5b4}

\bibitem[{{Howard} {et~al.}(2023){Howard}, {Kowalski}, {Flagg}, {MacGregor}, {Lim}, {Radica}, {Piaulet}, {Roy}, {Lafreni{\`e}re}, {Benneke}, {Brown}, {Espinoza}, {Doyon}, {Coulombe}, {Johnstone}, {Cowan}, {Jayawardhana}, {Turner}, \& {Dang}}]{howard2023}
{Howard}, W.~S., {Kowalski}, A.~F., {Flagg}, L., {et~al.} 2023, \apj, 959, 64, \dodoi{10.3847/1538-4357/acfe75}

\bibitem[{{Hunt-Walker} {et~al.}(2012){Hunt-Walker}, {Hilton}, {Kowalski}, {Hawley}, \& {Matthews}}]{huntwalker2012}
{Hunt-Walker}, N.~M., {Hilton}, E.~J., {Kowalski}, A.~F., {Hawley}, S.~L., \& {Matthews}, J.~M. 2012, \pasp, 124, 545, \dodoi{10.1086/666495}

\bibitem[{Hunter(2007)}]{Hunter:2007}
Hunter, J.~D. 2007, Computing in Science \& Engineering, 9, 90, \dodoi{10.1109/MCSE.2007.55}

\bibitem[{{Ilin} {et~al.}(2020){Ilin}, {Schmidt}, {Poppenh{\"a}ger}, {Davenport}, {Kristiansen}, \& {Omohundro}}]{ilin2020}
{Ilin}, E., {Schmidt}, S.~J., {Poppenh{\"a}ger}, K., {et~al.} 2020, arXiv e-prints, arXiv:2010.05576.
\newblock \doarXiv{2010.05576}

\bibitem[{Jenkins {et~al.}(2016)Jenkins, Twicken, McCauliff, Campbell, Sanderfer, Lung, Mansouri-Samani, Girouard, Tenenbaum, Klaus, Smith, Caldwell, Chacon, Henze, Heiges, Latham, Morgan, Swade, Rinehart, \& Vanderspek}]{jenkins_tess_2016}
Jenkins, J.~M., Twicken, J.~D., McCauliff, S., {et~al.} 2016, 9913, 99133E, \dodoi{10.1117/12.2233418}

\bibitem[{Jones {et~al.}(2001--)Jones, Oliphant, Peterson, {et~al.}}]{scipy}
Jones, E., Oliphant, T., Peterson, P., {et~al.} 2001--, {SciPy}: Open source scientific tools for {Python}.
\newblock \url{http://www.scipy.org/}

\bibitem[{Jönsson {et~al.}(2020)Jönsson, Holtzman, Allende~Prieto, Cunha, García-Hernández, Hasselquist, Masseron, Osorio, Shetrone, Smith, Stringfellow, Bizyaev, Edvardsson, Majewski, Mészáros, Souto, Zamora, Beaton, Bovy, Donor, Pinsonneault, Poovelil, \& Sobeck}]{jonsson_apogee_2020}
Jönsson, H., Holtzman, J.~A., Allende~Prieto, C., {et~al.} 2020, The Astronomical Journal, 160, 120, \dodoi{10.3847/1538-3881/aba592}

\bibitem[{Kluyver {et~al.}(2016)Kluyver, Ragan-Kelley, P{\'e}rez, Granger, Bussonnier, Frederic, Kelley, Hamrick, Grout, Corlay, {et~al.}}]{kluyver2016jupyter}
Kluyver, T., Ragan-Kelley, B., P{\'e}rez, F., {et~al.} 2016, in ELPUB, 87--90

\bibitem[{{Koch} {et~al.}(2010){Koch}, {Borucki}, {Basri}, {Batalha}, {Brown}, {Caldwell}, {Christensen-Dalsgaard}, {Cochran}, {DeVore}, {Dunham}, {Gautier}, {Geary}, {Gilliland}, {Gould}, {Jenkins}, {Kondo}, {Latham}, {Lissauer}, {Marcy}, {Monet}, {Sasselov}, {Boss}, {Brownlee}, {Caldwell}, {Dupree}, {Howell}, {Kjeldsen}, {Meibom}, {Morrison}, {Owen}, {Reitsema}, {Tarter}, {Bryson}, {Dotson}, {Gazis}, {Haas}, {Kolodziejczak}, {Rowe}, {Van Cleve}, {Allen}, {Chandrasekaran}, {Clarke}, {Li}, {Quintana}, {Tenenbaum}, {Twicken}, \& {Wu}}]{koch2010}
{Koch}, D.~G., {Borucki}, W.~J., {Basri}, G., {et~al.} 2010, \apjl, 713, L79, \dodoi{10.1088/2041-8205/713/2/L79}

\bibitem[{{Kopp} {et~al.}(2016){Kopp}, {Krivova}, {Wu}, \& {Lean}}]{kopp2016}
{Kopp}, G., {Krivova}, N., {Wu}, C.~J., \& {Lean}, J. 2016, \solphys, 291, 2951, \dodoi{10.1007/s11207-016-0853-x}

\bibitem[{{Kowalski} {et~al.}(2015){Kowalski}, {Hawley}, {Carlsson}, {Allred}, {Uitenbroek}, {Osten}, \& {Holman}}]{kowalski2015}
{Kowalski}, A.~F., {Hawley}, S.~L., {Carlsson}, M., {et~al.} 2015, \solphys, 290, 3487, \dodoi{10.1007/s11207-015-0708-x}

\bibitem[{{Kowalski} {et~al.}(2013){Kowalski}, {Hawley}, {Wisniewski}, {Osten}, {Hilton}, {Holtzman}, {Schmidt}, \& {Davenport}}]{kowalski2013}
{Kowalski}, A.~F., {Hawley}, S.~L., {Wisniewski}, J.~P., {et~al.} 2013, \apjs, 207, 15, \dodoi{10.1088/0067-0049/207/1/15}

\bibitem[{{Lacy} {et~al.}(1976){Lacy}, {Moffett}, \& {Evans}}]{lme1976}
{Lacy}, C.~H., {Moffett}, T.~J., \& {Evans}, D.~S. 1976, \apjs, 30, 85, \dodoi{10.1086/190358}

\bibitem[{{Lightkurve Collaboration} {et~al.}(2018){Lightkurve Collaboration}, {Cardoso}, {Hedges}, {Gully-Santiago}, {Saunders}, {Cody}, {Barclay}, {Hall}, {Sagear}, {Turtelboom}, {Zhang}, {Tzanidakis}, {Mighell}, {Coughlin}, {Bell}, {Berta-Thompson}, {Williams}, {Dotson}, \& {Barentsen}}]{2018ascl.soft12013L}
{Lightkurve Collaboration}, {Cardoso}, J.~V.~d.~M., {Hedges}, C., {et~al.} 2018, {Lightkurve: Kepler and TESS time series analysis in Python}, Astrophysics Source Code Library.
\newblock \doeprint{1812.013}

\bibitem[{{Lin} {et~al.}(2002){Lin}, {Dennis}, {Hurford}, {Smith}, {Zehnder}, {Harvey}, {Curtis}, {Pankow}, {Turin}, {Bester}, {Csillaghy}, {Lewis}, {Madden}, {van Beek}, {Appleby}, {Raudorf}, {McTiernan}, {Ramaty}, {Schmahl}, {Schwartz}, {Krucker}, {Abiad}, {Quinn}, {Berg}, {Hashii}, {Sterling}, {Jackson}, {Pratt}, {Campbell}, {Malone}, {Landis}, {Barrington-Leigh}, {Slassi-Sennou}, {Cork}, {Clark}, {Amato}, {Orwig}, {Boyle}, {Banks}, {Shirey}, {Tolbert}, {Zarro}, {Snow}, {Thomsen}, {Henneck}, {Mchedlishvili}, {Ming}, {Fivian}, {Jordan}, {Wanner}, {Crubb}, {Preble}, {Matranga}, {Benz}, {Hudson}, {Canfield}, {Holman}, {Crannell}, {Kosugi}, {Emslie}, {Vilmer}, {Brown}, {Johns-Krull}, {Aschwanden}, {Metcalf}, \& {Conway}}]{lin2002}
{Lin}, R.~P., {Dennis}, B.~R., {Hurford}, G.~J., {et~al.} 2002, \solphys, 210, 3, \dodoi{10.1023/A:1022428818870}

\bibitem[{{McQuillan} {et~al.}(2013){McQuillan}, {Aigrain}, \& {Mazeh}}]{mcquillan2013}
{McQuillan}, A., {Aigrain}, S., \& {Mazeh}, T. 2013, \mnras, 432, 1203, \dodoi{10.1093/mnras/stt536}

\bibitem[{{Montet} {et~al.}(2017){Montet}, {Tovar}, \& {Foreman-Mackey}}]{montet2017}
{Montet}, B.~T., {Tovar}, G., \& {Foreman-Mackey}, D. 2017, \apj, 851, 116, \dodoi{10.3847/1538-4357/aa9e00}

\bibitem[{{Morin} {et~al.}(2010){Morin}, {Donati}, {Petit}, {Delfosse}, {Forveille}, \& {Jardine}}]{morin2010}
{Morin}, J., {Donati}, J.-F., {Petit}, P., {et~al.} 2010, \mnras, 407, 2269, \dodoi{10.1111/j.1365-2966.2010.17101.x}

\bibitem[{Morris {et~al.}(2019)Morris, Davenport, Giles, Hebb, Hawley, Angus, Gilman, \& Agol}]{morris_solar_2019}
Morris, B.~M., Davenport, J. R.~A., Giles, H. A.~C., {et~al.} 2019, Monthly Notices of the Royal Astronomical Society, 484, 3244, \dodoi{10.1093/mnras/stz199}

\bibitem[{{Notsu} {et~al.}(2013){Notsu}, {Shibayama}, {Maehara}, {Notsu}, {Nagao}, {Honda}, {Ishii}, {Nogami}, \& {Shibata}}]{Notsu2013}
{Notsu}, Y., {Shibayama}, T., {Maehara}, H., {et~al.} 2013, \apj, 771, 127, \dodoi{10.1088/0004-637X/771/2/127}

\bibitem[{Oliphant(2007)}]{numpy}
Oliphant, T.~E. 2007, Computing in Science Engineering, 9, 10, \dodoi{10.1109/MCSE.2007.58}

\bibitem[{Oláh {et~al.}(2016)Oláh, Kővári, Petrovay, Soon, Baliunas, Kolláth, \& Vida}]{olah_magnetic_2016}
Oláh, K., Kővári, Z., Petrovay, K., {et~al.} 2016, Astronomy and Astrophysics, 590, A133, \dodoi{10.1051/0004-6361/201628479}

\bibitem[{pandas~development team(2024)}]{pandas_10697587}
pandas~development team, T. 2024, pandas-dev/pandas: Pandas, v2.2.1,  Zenodo, \dodoi{10.5281/zenodo.10697587}

\bibitem[{Pedregosa {et~al.}(2011)Pedregosa, Varoquaux, Gramfort, Michel, Thirion, Grisel, Blondel, Prettenhofer, Weiss, Dubourg, Vanderplas, Passos, Cournapeau, Brucher, Perrot, \& Duchesnay}]{scikit-learn}
Pedregosa, F., Varoquaux, G., Gramfort, A., {et~al.} 2011, Journal of Machine Learning Research, 12, 2825

\bibitem[{{Perez} \& {Granger}(2007)}]{2007CSE.....9c..21P}
{Perez}, F., \& {Granger}, B.~E. 2007, Computing in Science and Engineering, 9, 21, \dodoi{10.1109/MCSE.2007.53}

\bibitem[{Raetz \& Stelzer(2024)}]{raetz_long-term_2024}
Raetz, S., \& Stelzer, B. 2024, Long-term stellar activity of {M} dwarfs: {A} combined {K2} and {TESS} study of two early {M}-type stars,  arXiv, \dodoi{10.48550/arXiv.2404.16625}

\bibitem[{Ricker {et~al.}(2015)Ricker, Winn, Vanderspek, Latham, Bakos, Bean, Berta-Thompson, Brown, Buchhave, Butler, Butler, Chaplin, Charbonneau, Christensen-Dalsgaard, Clampin, Deming, Doty, De~Lee, Dressing, Dunham, Endl, Fressin, Ge, Henning, Holman, Howard, Ida, Jenkins, Jernigan, Johnson, Kaltenegger, Kawai, Kjeldsen, Laughlin, Levine, Lin, Lissauer, MacQueen, Marcy, McCullough, Morton, Narita, Paegert, Palle, Pepe, Pepper, Quirrenbach, Rinehart, Sasselov, Sato, Seager, Sozzetti, Stassun, Sullivan, Szentgyorgyi, Torres, Udry, \& Villasenor}]{ricker_transiting_2015}
Ricker, G.~R., Winn, J.~N., Vanderspek, R., {et~al.} 2015, Journal of Astronomical Telescopes, Instruments, and Systems, 1, 014003, \dodoi{10.1117/1.JATIS.1.1.014003}

\bibitem[{Schrijver \& Zwaan(2000)}]{schrijver_solar_2000}
Schrijver, C.~J., \& Zwaan, C. 2000, Solar and {Stellar} {Magnetic} {Activity}.
\newblock \url{https://ui.adsabs.harvard.edu/abs/2000ssma.book.....S}

\bibitem[{{Scoggins} {et~al.}(2019){Scoggins}, {Davenport}, \& {Covey}}]{scoggins2019}
{Scoggins}, M.~T., {Davenport}, J. R.~A., \& {Covey}, K.~R. 2019, Research Notes of the American Astronomical Society, 3, 137, \dodoi{10.3847/2515-5172/ab45a0}

\bibitem[{Shibayama {et~al.}(2013)Shibayama, Maehara, Notsu, Notsu, Nagao, Honda, Ishii, Nogami, \& Shibata}]{shibayama_superflares_2013}
Shibayama, T., Maehara, H., Notsu, S., {et~al.} 2013, The Astrophysical Journal Supplement Series, 209, 5, \dodoi{10.1088/0067-0049/209/1/5}

\bibitem[{Stassun {et~al.}(2018)Stassun, Oelkers, Pepper, Paegert, De~Lee, Torres, Latham, Charpinet, Dressing, Huber, Kane, Lépine, Mann, Muirhead, Rojas-Ayala, Silvotti, Fleming, Levine, \& Plavchan}]{stassun_tess_2018}
Stassun, K.~G., Oelkers, R.~J., Pepper, J., {et~al.} 2018, The Astronomical Journal, 156, 102, \dodoi{10.3847/1538-3881/aad050}

\bibitem[{Strassmeier(2005)}]{strassmeier_stellar_2005}
Strassmeier, K.~G. 2005, Astronomische Nachrichten, 326, 269, \dodoi{10.1002/asna.200410388}

\bibitem[{{Tovar Mendoza} {et~al.}(2022){Tovar Mendoza}, {Davenport}, {Agol}, {Jackman}, \& {Hawley}}]{tovar-mendoza2022}
{Tovar Mendoza}, G., {Davenport}, J. R.~A., {Agol}, E., {Jackman}, J. A.~G., \& {Hawley}, S.~L. 2022, arXiv e-prints, arXiv:2205.05706.
\newblock \doarXiv{2205.05706}

\bibitem[{{Usoskin}(2017)}]{usoskin2017}
{Usoskin}, I.~G. 2017, Living Reviews in Solar Physics, 14, 3, \dodoi{10.1007/s41116-017-0006-9}

\bibitem[{van~der Walt {et~al.}(2014)van~der Walt, {S}ch\"onberger, {Nunez-Iglesias}, {B}oulogne, {W}arner, {Y}ager, {G}ouillart, {Y}u, \& the scikit-image contributors}]{scikit-image}
van~der Walt, S., {S}ch\"onberger, J.~L., {Nunez-Iglesias}, J., {et~al.} 2014, PeerJ, 2, e453, \dodoi{10.7717/peerj.453}

\bibitem[{Van~Rossum \& Drake(2009)}]{python}
Van~Rossum, G., \& Drake, F.~L. 2009, Python 3 Reference Manual (Scotts Valley, CA: CreateSpace)

\bibitem[{Veronig {et~al.}(2002)Veronig, Temmer, Hanslmeier, Otruba, \& Messerotti}]{veronig_temporal_2002}
Veronig, A., Temmer, M., Hanslmeier, A., Otruba, W., \& Messerotti, M. 2002, Astronomy and Astrophysics, 382, 1070, \dodoi{10.1051/0004-6361:20011694}

\bibitem[{Virtanen {et~al.}(2020)Virtanen, Gommers, Oliphant, Haberland, Reddy, Cournapeau, Burovski, Peterson, Weckesser, Bright, {van der Walt}, Brett, Wilson, Millman, Mayorov, Nelson, Jones, Kern, Larson, Carey, Polat, Feng, Moore, {VanderPlas}, Laxalde, Perktold, Cimrman, Henriksen, Quintero, Harris, Archibald, Ribeiro, Pedregosa, {van Mulbregt}, \& {SciPy 1.0 Contributors}}]{2020SciPy-NMeth}
Virtanen, P., Gommers, R., Oliphant, T.~E., {et~al.} 2020, Nature Methods, 17, 261, \dodoi{10.1038/s41592-019-0686-2}

\bibitem[{Wagg \& Broekgaarden(2024)}]{software-citation-station-zenodo}
Wagg, T., \& Broekgaarden, F. 2024, The Software Citation Station,  Zenodo, \dodoi{10.5281/zenodo.11292917}

\bibitem[{{Wagg} \& {Broekgaarden}(2024)}]{software-citation-station-paper}
{Wagg}, T., \& {Broekgaarden}, F.~S. 2024, arXiv e-prints, arXiv:2406.04405.
\newblock \doarXiv{2406.04405}

\bibitem[{Wainer {et~al.}(2022)Wainer, Johnson, Seth, Torresvillanueva, Dalcanton, Durbin, Dolphin, Weisz, Williams, \& {Phatter Collaboration}}]{wainer_panchromatic_2022}
Wainer, T.~M., Johnson, L.~C., Seth, A.~C., {et~al.} 2022, The Astrophysical Journal, 928, 15, \dodoi{10.3847/1538-4357/ac51cf}

\bibitem[{Wang {et~al.}(2005)Wang, Lean, \& Sheeley}]{wang_modeling_2005}
Wang, Y.~M., Lean, J.~L., \& Sheeley, Jr., N.~R. 2005, The Astrophysical Journal, 625, 522, \dodoi{10.1086/429689}

\bibitem[{{W}es {M}c{K}inney(2010)}]{mckinney-proc-scipy-2010}
{W}es {M}c{K}inney. 2010, in {P}roceedings of the 9th {P}ython in {S}cience {C}onference, ed. {S}t\'efan van~der {W}alt \& {J}arrod {M}illman, 56 -- 61, \dodoi{10.25080/Majora-92bf1922-00a}

\bibitem[{{Yan} {et~al.}(2012){Yan}, {Deng}, {Qu}, {Xu}, \& {Kong}}]{yan2012}
{Yan}, X.~L., {Deng}, L.~H., {Qu}, Z.~Q., {Xu}, C.~L., \& {Kong}, D.~F. 2012, Journal of Astrophysics and Astronomy, 33, 387, \dodoi{10.1007/s12036-012-9153-5}

\bibitem[{Zhu {et~al.}(2015)Zhu, Liu, Alexander, Sun, \& McAteer}]{zhu_complex_2015}
Zhu, C., Liu, R., Alexander, D., Sun, X., \& McAteer, R. T.~J. 2015, The Astrophysical Journal, 813, 60, \dodoi{10.1088/0004-637X/813/1/60}

\end{thebibliography}

\end{document}